\documentclass[11pt,a4paper]{article} 
\usepackage{epsfig}
\usepackage{graphicx,psfrag}
\usepackage{multirow}
\usepackage{slashed}
\usepackage{pstricks}
\usepackage{caption}
\usepackage{subcaption}
\usepackage{url}
\usepackage{braket}
\usepackage{jheppub}
\usepackage{enumerate}
\usepackage{wasysym} 
\usepackage{mathrsfs} 
\usepackage{amsfonts} 
\usepackage{amsbsy} 
\usepackage{amscd}
\usepackage{color}

\makeatletter

\def\eeq{\end{equation}}
\def\beq{\begin{equation}}

\newcommand{\Rmnum}[1]{\expandafter\@slowromancap\romannumeral #1@}
\makeatother

\title{Extended scalar sectors, effective operators and observed data}

\author[a]{Atri Dey,}
\author[a]{Jayita Lahiri,} 
\author[a]{Biswarup Mukhopadhyaya} 
  \affiliation[a]{Regional Centre for Accelerator-based Particle Physics,
Harish-Chandra Research Institute, HBNI,
Chhatnag Road, Jhunsi, Allahabad - 211 019, India} 
\emailAdd{atridey@hri.res.in}
\emailAdd{jayitalahiri@hri.res.in}
\emailAdd{biswarup@hri.res.in}

\abstract{The available data on the 125 GeV scalar $h$ is analysed to explore the room for new physics in
the electroweak symmetry breaking sector. The first part of the study is model-independent,
with $h$ couplings to standard model particles scaled by quantities that are taken to be free parameters. At the same time, the additional loop contributions to $h \rightarrow \gamma\gamma$
and $h \rightarrow Z\gamma$, mediated by charged scalar contributions in the extended
scalar sector,  are treated in terms of gauge-invariant effective operators. Having justified
this approach for cases where the concerned scalar masses are a little above the $Z$-boson mass,
we fit the existing data to obtain marginalized 1$\sigma$ and 2$\sigma$ regions in the
space of the coefficients of such effective operators, where the limit on the $h \rightarrow Z\gamma$ branching ratio is used as a constraint. The correlation between, say, the
gluon fusion and vector-boson fusion channels, as reflected in a non-diagonal
covariance matrix, is taken into account. After thus obtaining model-independent
fits, the allowed values of the coefficients are translated into permissible regions
of the parameter spaces of several specific models. In this spirit we constrain four different
types of two Higgs doublet models, and also models with one or two $Y = 2$ scalar triplets,
taking into account the correlatedness of the scale factors in $h$-interactions and the
various couplings of charged Higgs states in each extended scenario.}

\preprint{HRI-RECAPP-2018-008\\$\textrm{}$}

\begin{document}

\maketitle

\newpage

\section{Introduction}
\label{intro}

The ATLAS and CMS experiments at the Large Hadron Collider
(LHC)~\cite{Khachatryan:2016vau} have discovered a new boson mass 125
GeV. Though the properties of this particle are similar to those of
the Higgs boson predicted in the standard model (SM) of electroweak
interactions, everyone is on the look-out for any small difference
that may reveal the participation of some new physics.  It is thus
imperative to closely examine all interactions (including supposedly
`effective' ones) of this particle with SM fermions and gauge
bosons. The accumulation of data has helped a lot in pinning down
uncertainties here; scopes of departure, however, still
remain. Besides, issues ranging from naturalness of the Higgs mass to
the dark matter content of our universe continue to provide impetus
for physics beyond the standard model (BSM). Theoretical models
extending the SM, including those augmenting the electroweak symmetry
breaking sector, are thus explored. It is a natural endeavour to link
the contributions of such new physics, in the form of modified Higgs
interaction strengths as well as effective operators generated by
them, to the departure from unity in the Higgs signal strengths in
various final states $f$, defined as $\mu^f =
{\frac{\sigma^f}{\sigma^f_{SM}}}$. Fitting the available data with 
various $\mu^f$, therefore, enables one to analyze allowed strengths
of effective operators generated in various models and ultimately
constrain the model parameters themselves. This is particularly useful
if the new particles belonging to any new physics scenario do not have
a copious rate for direct production, but show up (i) by participating
off-shell, and leading to higher-dimensional effective operators, and
(ii) by modifying the coupling strength(s) of the 125 GeV scalar to
the other SM particles. The present study is devoted to such a
situation.

Interestingly, significant constraints arise from the signal strengths for 
the loop induced decay channel such as $h \rightarrow \gamma \gamma$.
$h \rightarrow Z \gamma$, although yet unobserved, can also provide
strong limits on new physics contributions, especially with accumulating 
luminosity. These limits can be coupled with those  arising from
tree-level decay modes such as $h \rightarrow WW, ZZ$ etc. where perceptible
effects can come mostly via scaling of the SM coupling by a factor $\kappa$.

The effect of high scale physics on low-energy processes can be
formulated in terms of higher dimension operators in the Lagrangian,
which will be suppressed by the new physics scale $\Lambda$. These
higher dimensional operators can be derived from an
SU(2)$_L\times$U(1)$_Y$-invariant basis, as they result from physics
corresponding to a scale much higher than the electroweak scale. It is
not possible to construct dimension-5 operators of this kind.
Therefore we concentrate on dimension-6 gauge-invariant operators.  As
mentioned above, we mainly focus on two loop-induced decays, namely,
$h \rightarrow \gamma \gamma$ and $h \rightarrow Z \gamma$. Thus the
most general dimension-6 operators which give rise to $hVV$ vertices
($V = W, Z$) get constrained from the rates of these processes,
as tree-level processes are unlikely to bear clear enough stamps of
of such operators. Constraints on these higher
dimensional operators as well as the scale factors $\kappa$ have been
extensively studied on the basis of electroweak precision test and
global fits of Higgs data in
~\cite{Banerjee:2012xc,Masso:2012eq,Corbett:2012ja,
  Falkowski:2013dza,Corbett:2013pja,Dumont:2013wma,
  Gainer:2013rxa,Corbett:2013hia,Elias-Miro:2013mua,Einhorn:2013tja,
  Pomarol:2013zra,Banerjee:2013apa}.

Here we have adopted a slightly different formulation of new physics
contributions to the aforementioned loop-induced decays. The
consequences of new physics have been divided into two categories. The
first of these is the scaling of the couplings
$ht\bar{t}, hb\bar{b}, h\tau\bar{\tau}, hVV$, where modifications to the
SM couplings are inevitable when additional fields mixing with the
ones in SM are present. As we have already mentioned, such
modifications usually override the effects of higher-dimensional
operators. Such scaling also affects loop effects such as
$h\rightarrow \gamma\gamma, Z\gamma$, via modified vertices in the loop
where, W's or top quarks are involved, and the corresponding amplitudes
can be written down in terms of the scale factors $\kappa$. The
second category consists in loop diagrams mediated by new particles
such as charged scalars. Their contributions to loop amplitudes, we
argue, can be expressed in isolation in terms of effective couplings. 
As shown in section~\ref{model} , these couplings, at least
those ensuing from additional scalar fields,
can be treated as ones derived by the aforesaid gauge-invariant dimension-6
operators, so long as the masses of the new particles are gauge-invariant
and at least a little above the electroweak symmetry breaking (EWSB) scale.

Keeping this in mind, we perform a global fit of the currently
available Higgs data to constrain the full parameter space including
scale factors and the Wilson coefficients corresponding to various
dimension-6 operators. While using the Higgs signal strength data in
various channels we take into account the correlation between various
production processes such as gluon fusion and vector-boson fusion,
thus including non-diagonal covariant matrices in our analysis.
Model-independent 2$\sigma$ regions for the
$\kappa$'s well as the Wilson coefficients are thus obtained.

We select specific models in the next step of the analysis. The present 
study is restricted to additional scalars, and includes various kinds
of two-Higgs doublet models (2HDM) as well as those with one and two
scalar triplets of the kind introduced in Type-II see-saw mechanism.
The marginalized 2$\sigma$ regions in the space of dimension-6 operators 
are then recast, keeping track of the correlation
between the scale factors $\kappa$ and the Wilson coefficients
for each model. These are finally translated into constraints in the space of
masses and coupling strengths pertaining to all the models.
% {\bf some clarification for 2HDM is required}. 

 There is as of now just an upper limit on the signal strength in the
 $Z\gamma$ channel. Keeping this in mind, We further construct a ratio
 involving the signal strengths for $Z\gamma$ and $\gamma\gamma$,
 which can help us distinguish between various new physics models.  If
 a charged scalar (which can in principle contribute to both the above
 final states) is discovered later, this ratio will help us in
 narrowing down possibilities regarding the SU(2) multiplet it could
 be a part of. In case of a future discovery of one or more heavy
 Higgs, this can enable one to distinguish among various theories with
 extended electroweak symmetry breaking sectors, following the
 strategy developed in this paper.

 In a nutshell, the novel features in our
 study are the following:

\begin{itemize}
\item We have segregated two types of modification, namely,
   those in tree-level couplings of the Higgs to SM particles, and
   the contribution of loop integrals involving BSM mediators
   to processes disallowed at the tree-level.

\item The latter contribution mentioned above has been parametrized
  in isolation in terms of dimension-6 operators, and the legitimacy
  of doing this has been demonstrated.

\item The effect of the former modification to tree-level as well as
  loop-induced decays have been taken into account, while the
  contribution of dimension-6 operators have been assumed, as a
  pragmatic measure, to be confined to loop-induced processes.

\item The effective operators has been shown to work for just a subset
  of the one-loop diagrams contributing to effective $h\gamma\gamma$
  and $hZ\gamma$ couplings, even when the masses of new scalars
  running in these loops are not much above the electroweak scale.

\item Non-diagonal covariance matrices have been used.

\item We then establish a connection between allowed parameter regions
  in the model-independent approach with those of specific new physics
  models.

\item Finally we propose using the ratio of rates in the
  $\gamma\gamma$ and $Z\gamma$ channels, potentially observable at
  the LHC with accumulating luminosity. This allows disentanglement of
  various types of new physics. 
\end{itemize}

The paper is organized as follows. In Section~\ref{np}, we discuss the
parametrization of modified Higgs couplings involving both scaling
and higher-dimensional operators. A global fit carried out on an
8-dimensional parameter space, taking into account the correlation
among various production channels of the Higgs particle, is reported in
Section~\ref{fit} . Section~\ref{model} contains a discussion of
several extended Higgs models, wherefrom we arrive at the resulting constraints 
on the corresponding models. We summarize and
conclude in Section~\ref{summary}. Some expressions relevant to the loop amplitudes in $h \rightarrow \gamma \gamma$ and $h \rightarrow Z \gamma$ are listed in the Appendices.

\section{Modified Higgs couplings}
\label{np}

As mentioned in the previous section, there can be an extended Higgs
sector comprising of additional neutral and charged scalars. Their
mixing may cause the coupling of the 125 GeV scalar to SM particles to
be modified \footnote{In practice, similar modifications may arise due
  to the presence of additional fermions and gauge bosons, having
  mixing with the SM particles. We specifically mention additional
  scalars because the model-dependent part of our analysis is confined
  to such scenarios only.}.  Modification may generally occur in two ways.
First, there can be scaling of the Higgs couplings, with unaltered Lorentz
structure of the corresponding vertices. 
Such scaling can be expressed as

\begin{eqnarray}
{\tilde{g}}_{hVV} &=& {\kappa_v} \times g_{hVV} \\  
{\tilde{g}}_{ht\bar t} &=& {\kappa_t} \times g_{ht\bar t}  \\  
{\tilde{g}}_{hb\bar b} &=& {\kappa_b} \times g_{hb \bar b}  \\  
{\tilde{g}}_{h\tau \bar{\tau}} &=& {\kappa_{\tau}} \times g_{h\tau \bar{\tau}}
\end{eqnarray}
Where $g_{hVV}, g_{ht\bar t}, g_{hb \bar b}$ and $g_{h\tau \bar{\tau}}$ are the couplings of the Higgs to the gauge bosons and the fermions in the SM. 
The couplings of Higgs to $W$
boson and $Z$ boson are scaled in the same way here. 
This is because the custodial SU(2) symmetry, ensuring tree-level unity of
the $\rho$-parameter, is otherwise at stake, unless additional
contributions to $\rho$ are built into the theory.

Moreover, there can be
heavy states running in the loop modifying Higgs
couplings. A general approach to parametrize such
modification is to express it in
terms of gauge-invariant higher dimensional effective operators. Here one
normally expects that the new physics  here is at a high scale (at
least a TeV or thereabout), thus justifying  SU$(2)_L\times$
U$(1)_Y$ invariance of the operator(s) involved. However, one can 
see that such TeV-scale suppressants may arise even for lower masses
in the loop, thanks to the factors ${\cal O}$$\sim 16\pi^2$ in the loop integrals.
Gauge invariance of the Wilson coefficients, will, require that
the masses of particles running in the loops should arise from
SU$(2)_L\times$ U$(1)_Y$ invariant terms. This requirement is satisfied
in each model, involving extension of the electroweak symmetry breaking
sector, used by us for illustration.

All Higgs interactions should in principle be modified via such
operators. However, couplings which exist at the SM at tree-level,
namely, $hWW$, $hZZ$, $ht\bar t$, $hb\bar b$, $h \tau \bar{\tau}$, are
rather nominally affected by higher-dimensional terms with (at least)
TeV-scale suppression. Thus one can, for all practical purpose,
neglect such modification compared to the tree-level couplings and the
scaling effects, if any. On the other hand, the $h\gamma\gamma$ and
$hZ\gamma$ vertices which appear only at the one-loop level are
expected to have relatively non-negligible contribution from the
additional diagrams with new particles running in the loops.

Dimension-6 effective interactions involving a Higgs and two gauge bosons
can be expressed in terms of the following gauge-invariant
operators~\cite{Hagiwara:1992eh,Hagiwara:1993ck}:

\begin{eqnarray}
{\cal O}_{BB} &=& \frac{f_{BB}}{\Lambda^2}{\Phi}^{\dagger}\hat{B}_{\mu \nu}\hat{B}^{\mu \nu}\Phi \nonumber \\
{\cal O}_{WW} &=& \frac{f_{WW}}{\Lambda^2}{\Phi}^{\dagger}\hat{W}_{\mu \nu}\hat{W}^{\mu \nu}\Phi  \nonumber \\
{\cal O}_{B} &=& \frac{f_{B}}{\Lambda^2}{D_{\mu}\Phi}^{\dagger}\hat{B}^{\mu \nu}D_{\nu}\Phi  \nonumber \\
{\cal O}_{W} &=& \frac{f_{W}}{\Lambda^2}{D_{\mu}\Phi}^{\dagger}\hat{W}^{\mu \nu}D_{\nu}\Phi 
\label{eft} 
\end{eqnarray}
Where

\begin{equation}
B_{\mu \nu} = \partial_{\mu}B_{\nu} - \partial_{\nu}B_{\mu}
\end{equation}
\begin{equation}
\hat{B}_{\mu \nu} = i\frac{g\tan\theta_w}{2} B_{\mu \nu}
\end{equation}

and

\begin{equation}
{W^a}_{\mu \nu} = \partial_{\mu}{W^a}_{\nu} -
\partial_{\nu}{W^a}_{\mu}-g \epsilon^{abc} {W^b}_{\mu} {W^c}_{\mu}
\end{equation}
\begin{equation}
\hat{W}_{\mu \nu} = i\frac{g}{2} \sigma^a {W^a}_{\mu \nu}
\end{equation}

These constitute the most general set of dimension-6 effective
operators which give rise to the $h\gamma\gamma$ and $hZ\gamma$
vertices. ${\cal O}_{B}$ and ${\cal O}_{W}$ contribute only to the
$hZ\gamma$ vertex but not to $h\gamma\gamma$, since the component of
Higgs doublet that acquires a vacuum expectation value (vev) has no
electromagnetic charge. Furthermore, there is the operator ${\cal
  O}_{BW}$ which is of the form
$\frac{f_{BW}}{\Lambda^2}{\Phi}^{\dagger}\hat{B}_{\mu
  \nu}\hat{W}^{\mu\nu}\Phi$.  However, it leads to tree-level mixing
between $\gamma$ and $Z$, thereby altering the $Z$-mass eigenstate,
while keeping the $W$ mass eigenstate unaffected. Consequently, this
term breaks the custodial SU(2) and therefore must be highly
suppressed in general.  This operator is not not considered in the
present analysis.

The contribution of the operators ${\cal O}_{BB}$, ${\cal O}_{WW}$,
${\cal O}_{B}$ and ${\cal O}_{W}$ to the $h\gamma \gamma$ and $h Z
\gamma$ vertices can be obtained from the Equation~\ref{eft}. However, as has been
already mentioned, we have chosen to express only that part of the contributions
to the loop-induced amplitude in each case, which is the consequence of new 
particles running in the loops. The part coming from loops
induced by W and the standard model fermions are used in their
already available forms, with appropriate scaling of couplings
denoted by the various $\kappa$-factors. Such a description in terms
of gauge-invariant effective operators is possible, especially if the Wilson
coefficients generated by the BSM loops are functions of masses that
do not break electroweak gauge invariance. Here we apply this method
to scenarios presumed to have additional scalars only, whose mass terms
are gauge-invariant to start with and more will be said on this in Section 4.
\footnote{This analysis is legitimate at the one-loop level, where there is no
interplay of SM and BSM loops. Although the effective operators
in principle encapsulate contribution at all orders, it has been assumed that
the dominant contribution comes at the one-loop level. Thus the two parts,
namely those from the scaled SM loops and the ones from new particle 
contributions, can be taken to be of the same order of perturbation.}

\begin{figure}
\centering
\includegraphics[width=12.8cm, height=9.3cm]{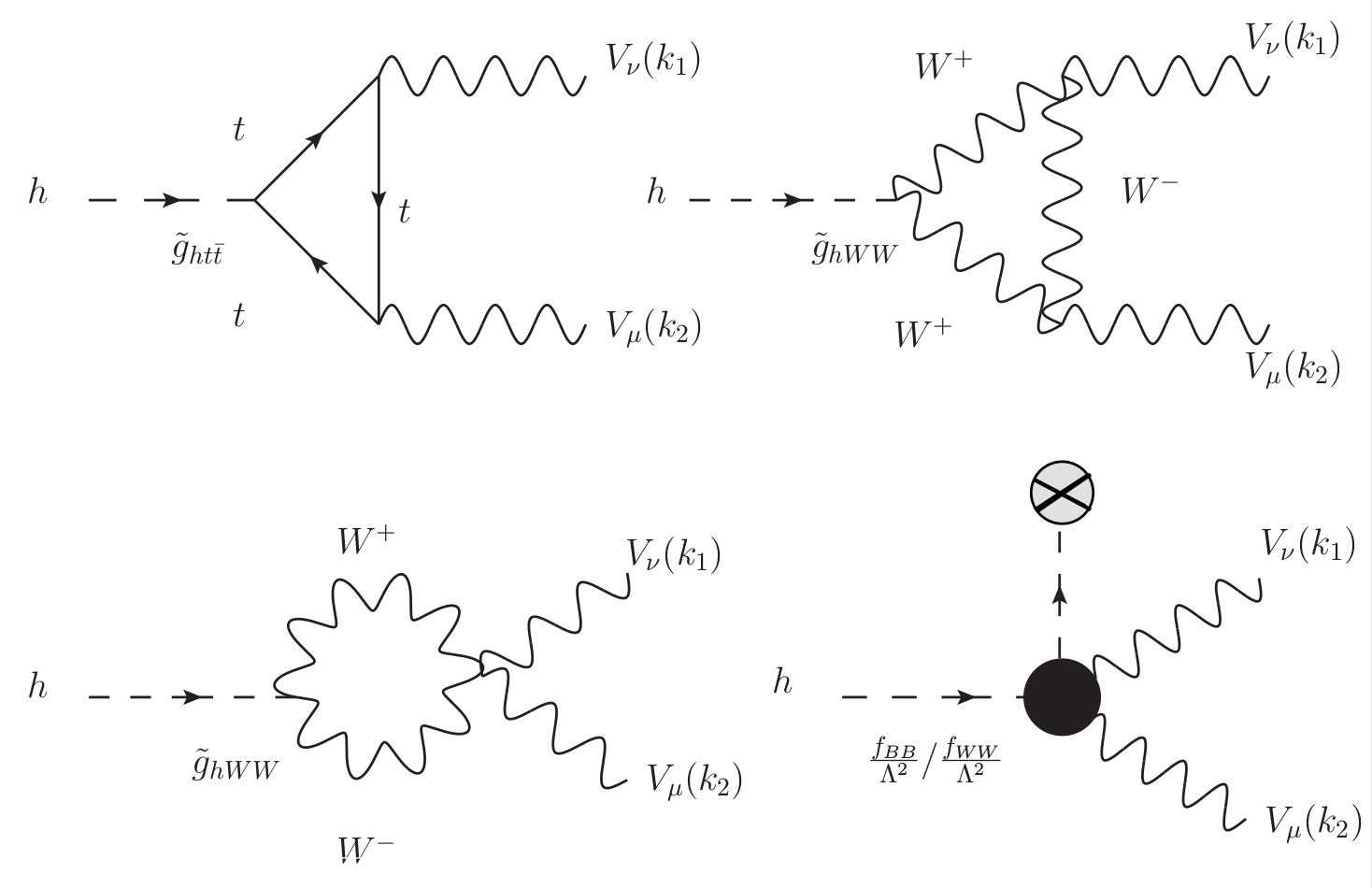}
\caption{Feynman diagrams for $h \rightarrow \gamma \gamma$
  and $h \rightarrow Z \gamma$ in the most general situation. Contributions mediated by fields other than those in SM are lumped in the blob.}
\label{figure1}
\end{figure}

The parts of the contributions from the effective operators 
${\cal O}_{BB}$ and ${\cal O}_{WW}$ are given in
Table.~\ref{obbww} while similar contributions from ${\cal O}_{B}$ and ${\cal
  O}_{W}$ are given in Table.~\ref{obw}.
\footnote{We proved on the basis of the ansatz that the contribution
  to $h \rightarrow \gamma \gamma$ and $h \rightarrow Z \gamma$ from
  loops comprising new particles, which are interrelated by $SU(2)_{L}
  \times U(1)_{Y}$, are expressible as shown in the tables. This is
  possible only if the Wilson coefficients are insensitive to $m_{Z}$
  and therefore the EWSB scale. This ansatz
  is justified in Section 4.}  We show in Figure~\ref{figure1}, the
Feynman diagrams contributing to the decays $h\rightarrow \gamma
\gamma$ and $h \rightarrow Z \gamma$, including those with scaled
contribution of SM and and those expressed in terms of loop induced
dimension-6 operators.

\begin{table}[!h]

\begin{center}
\begin{tabular}{|c||c|c|} \hline
%%%%%% Title row starts here
& $h \rightarrow \gamma \gamma$ & $h \rightarrow Z \gamma$ \\ \hline\hline
%%%%%% Row Foo starts here
${\cal O}_{BB}$ &
\begin{tabular}{c} $ -i{\cal M}_{BB} = $\\ $4\frac{f_{BB}}{\Lambda^2}\frac{g^2}{4}\sin^{2}\theta_{w}v\times$\\\footnotesize{$(k_1.k_2 g_{\mu \nu} - k_{1\mu}k_{2\nu})\epsilon^{*\mu}(k_2)\epsilon^{*\nu}(k_1) $} \\
\end{tabular} &
\begin{tabular}{c} $ -i{\cal M}_{BB} = $\\ $-4\frac{f_{BB}}{\Lambda^2}\frac{g^2}{2}\frac{\sin^{3}\theta_{w}}{\cos\theta_{w}}v\times$\\\footnotesize{$(k_1.k_2 g_{\mu \nu} - k_{1\mu}k_{2\nu})\epsilon^{*\mu}(k_2)\epsilon^{*\nu}(k_1) $ }\\
\end{tabular} \\ \hline
%%%%%% Row Bar starts here
${\cal O}_{WW}$ &
\begin{tabular}{c}  $ -i{\cal M}_{WW} = $\\ $4\frac{f_{WW}}{\Lambda^2}\frac{g^2}{4}\sin^{2}\theta_{w}v\times$\\\footnotesize{$(k_1.k_2 g_{\mu \nu} - k_{1\mu}k_{2\nu})\epsilon^{*\mu}(k_2)\epsilon^{*\nu}(k_1) $} \\ \\
\end{tabular} &
\begin{tabular}{c} $ -i{\cal M}_{WW} = $\\ $4\frac{f_{WW}}{\Lambda^2}\frac{g^2}{2}\sin\theta_{w}\cos\theta_{w}v\times$\\\footnotesize{$(k_1.k_2 g_{\mu \nu} - k_{1\mu}k_{2\nu})\epsilon^{*\mu}(k_2)\epsilon^{*\nu}(k_1) $ }\\ 
\end{tabular} \\ \hline
\end{tabular}
\end{center}
\caption{The part of the amplitudes for 
$h \rightarrow \gamma \gamma$ and $h \rightarrow Z \gamma$ 
coming from new particle loops, and expressed in terms of the dimension-6 operators 
${\cal O}_{BB}$ and ${\cal O}_{WW}$.}
\label{obbww}
\end{table}

\begin{table}[!h]

\begin{center}

\begin{tabular}{|c||c|c|} \hline
%%%%%% Title row starts here
& $h \rightarrow \gamma \gamma$ & $h \rightarrow Z \gamma$ \\ \hline\hline
%%%%%% Row Foo starts here
${\cal O}_{B}$ &
\begin{tabular}{c} 0 \\
\end{tabular} &
\begin{tabular}{c} $ -i{\cal M}_{B} = $\\ $-2\frac{f_{B}}{\Lambda^2}\frac{gg_{z}}{2}\sin\theta_{w}v\times$\\\small{$(k_1.k_2 g_{\mu \nu} - k_{1\mu}k_{2\nu})\epsilon^{*\mu}(k_2)\epsilon^{*\nu}(k_1) $ }\\
\end{tabular} \\ \hline
%%%%%% Row Bar starts here

%%%%%%%%%%%%%%
${\cal O}_{W}$ &
\begin{tabular}{c}  0 \\ 
\end{tabular} &
\begin{tabular}{c} $ -i{\cal M}_{W} = $\\ $2\frac{f_{W}}{\Lambda^2}\frac{gg_{z}}{4}\sin\theta_{w}v\times$\\\small{$(k_1.k_2 g_{\mu \nu} - k_{1\mu}k_{2\nu})\epsilon^{*\mu}(k_2)\epsilon^{*\nu}(k_1) $ }\\ 
\end{tabular} \\ \hline
\end{tabular}
%\caption{gvhj}
\end{center}
\caption{The part of the amplitudes for 
$h \rightarrow Z \gamma$  coming from new particle loops, and expressed
in terms of the dimension-6 operators ${\cal O}_{B}$ and ${\cal O}_{W}$.}
\label{obw}
\end{table}

Rather precise results on $h \rightarrow \gamma\gamma$ are already
available, and thus the disentanglement of the scaling effect and the
higher-dimensional contributions is expected using this channel in a
global fit. Similar results on $h \rightarrow Z\gamma$, dependent on
leptonic decays of $Z$, are awaited till more luminosity
accumulates. However, the upper limit on its production $\times$
branching ratio~\cite{Aaboud:2017uhw} available at any stage has served as a useful
constraining factor, when it comes to obtaining the statistically
favored ranges of scaling factors as well as Wilson
coefficients. This will be discussed in greater detail in Section 4.

The role and usefulness of higher-dimensional operators in dealing
with unknown physics in Higgs phenomenology have already been
discussed in a large body of works found in the literature
~\cite{Banerjee:2012xc,Masso:2012eq,Corbett:2012ja,Falkowski:2013dza,
 Corbett:2013pja,Dumont:2013wma,Gainer:2013rxa,
Corbett:2013hia,Elias-Miro:2013mua,
 Einhorn:2013tja,Pomarol:2013zra,Banerjee:2013apa,Banerjee:2015bla}. 
For example, in~\cite{Banerjee:2015bla}, ways towards improved 
understanding on the dimension-6
$hVV$ operators in the high luminosity run of the LHC have been
suggested. The utility of various ratios of signal strengths 
in different channels has figured in discussions found there. One 
such ratio, namely that of the signal strengths in the $\gamma\gamma$ and
$Z\gamma$ channels, has been adopted as a component of the present analysis.

\section{The global fit}
\label{fit}

After parametrizing new physics effects in the manner discussed above,
we investigate the region of parameter space favored by the 8 and 13
TeV results at the LHC. Our parameter space is eight-dimensional,
spanned by the four scale factors $\kappa_V$, $\kappa_t$, $\kappa_b$
and $\kappa_{\tau}$ (which parametrize the modification of SM
tree-level $hVV, ht \bar t, h b \bar b, h \tau \bar{\tau}$ couplings)
and $f_{BB}, f_{WW}, f_B$ and $f_W$, which are the Wilson coefficients
in the dimension-6 $hVV$ operators. In order to constrain new physics
from experimental data, one needs to construct the likelihood
function.  This in principle is a non-trivial task, as experiment only
provides the best fit and the $1 \sigma$ interval, but not the full
likelihood function. However, assuming the measurements to be
Gaussian, the Log-likelihood can be written as

\begin{equation}
-2\text{log}L(\mu) =\left(\frac{\mu - \hat\mu}{\Delta \mu}\right)^2
\end{equation}

Where $\hat {\mu} $ is the experimental best fit value of the measured
signal strength in some channel and $\Delta\mu$ is the corresponding
standard deviation. If there are more than one independent
measurements, then the full likelihood is a product of individual
likelihoods. Consequently the combined $\chi^2$ is 

\begin{equation}
L(\mu) = \prod_{i=1}^{n} L_{i}(\mu) \quad \Rightarrow \quad
\chi^2(\mu) = \sum_{i=1}^{n} \chi_i^2(\mu) = \sum_{i=1}^{n}
\left(\frac{\mu - \hat\mu_i}{\Delta \mu_i}\right)^2
\,. \label{eq:likesimpleprod}
\end{equation}

We would like to emphasize that the combined $\chi^2$
computation will follow Equation~\ref{eq:likesimpleprod} only when all the
experimental measurements denoted by subscript $i$, are independent of
each other. But in reality that assumption does not always
hold. Various experimental searches share same systematic uncertainties 
and thus some
correlation may exist between them. For example, different production channels 
leading to the same final can  develop
such correlation owing to misidentification of the production
processes. Under the Gaussian approximation these correlations affect
the Log-likelihood function in the following manner.

\begin{equation}
-2 \log L(\mu) = \chi^2(\mu) = 
(\mu - \hat{\mu}_i)^T C^{-1}_{ij} (\mu - \hat{\mu}_j) \,, 
\label{eq:gausscovmatrix}
\end{equation}

Where $C^{-1}$ is the inverse of the covariance matrix $C_{ij}$ =
cov($\hat{\mu}_i,\hat{\mu}_j)$ In the case where the measurements are
independent, the covariance matrix is diagonal and $C^{-1}_{ii}$ will
denote the variance $\sigma_i^2$. The correlations between gluon
fusion and vector-boson fusion production for each of the major Higgs
decay channels are found in~\cite{Khachatryan:2016vau,CMS:2017rli,CMS:2017jkd,CMS:2017pzi,Sirunyan:2017khh,ATLAS-CONF-2017-045,ATLAS-CONF-2017-043}.  We
extract the elements of the covariance matrices from those ellipses
and calculate the combined $\chi^2$ for each point of the parameter
space. This $\chi^2$ is then minimized with the help of the package
MCMC~\cite{ForemanMackey:2012ig}, wherefrom one obtains the region
allowed by the experimental data at the 1-and $2 \sigma$ levels.

\begin{figure*}[!hptb]
\centering
\includegraphics[width=5.9cm, height=4.9cm]{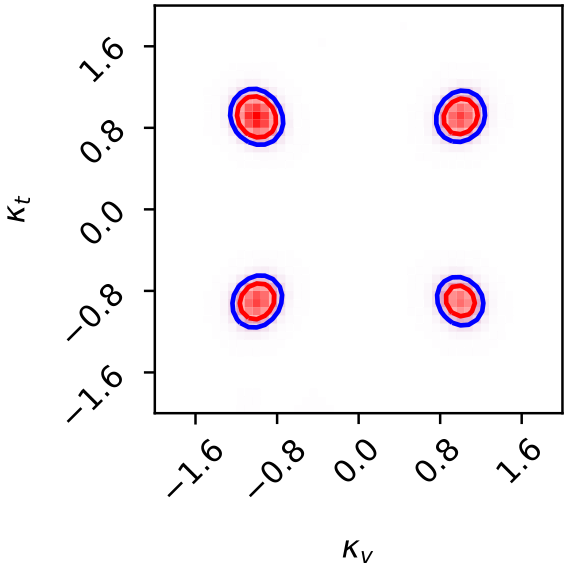}\hspace*{0.8cm}
\includegraphics[width=5.9cm, height=4.9cm]{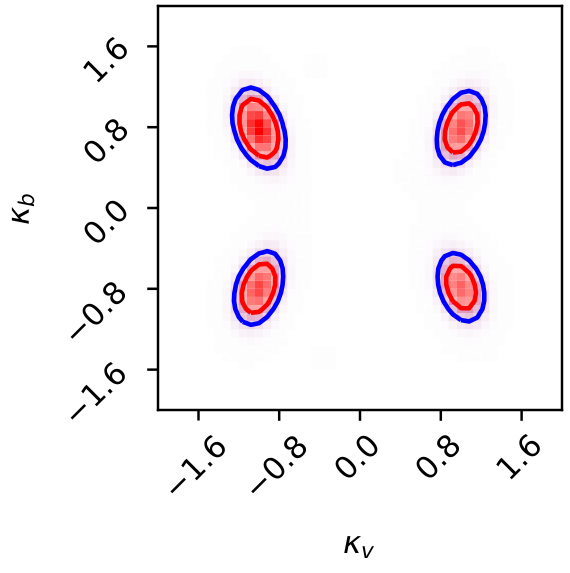}
\includegraphics[width=5.9cm, height=4.9cm]{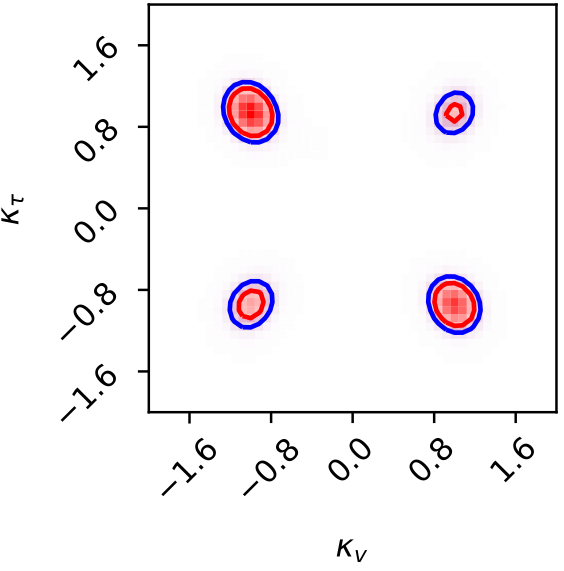}\hspace*{0.8cm}
\includegraphics[width=5.9cm, height=4.9cm]{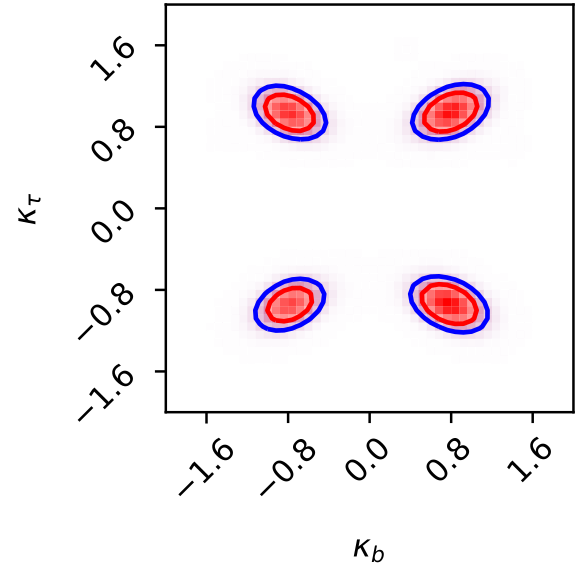}
\includegraphics[width=5.9cm, height=4.9cm]{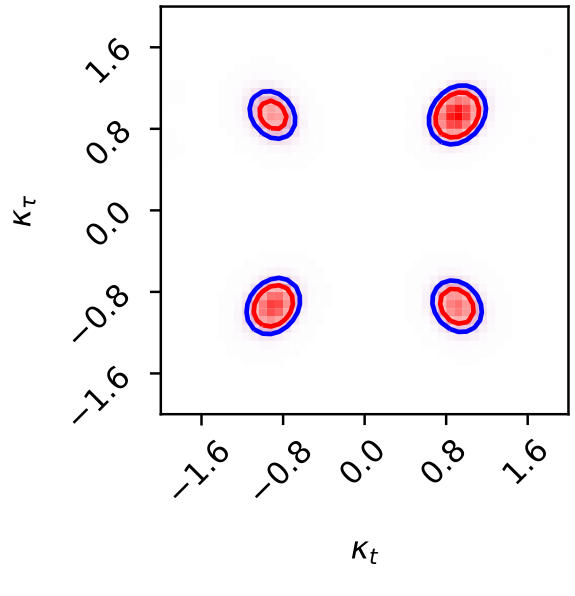}\hspace*{0.8cm}
\includegraphics[width=5.9cm, height=4.9cm]{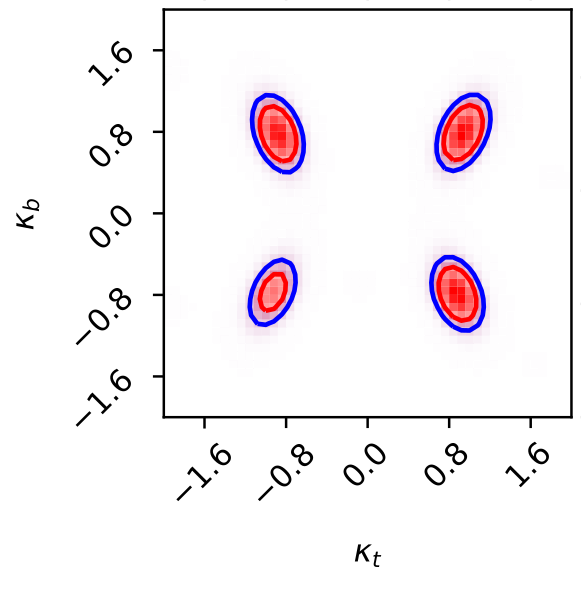}
\caption{Allowed regions at $1\sigma$(red) and $2\sigma$(blue)
  levels in the parameter space of scale factors}
\label{figure2}
\end{figure*}

\begin{figure*}[!hptb]
\centering
\includegraphics[width=5.9cm, height=4.9cm]{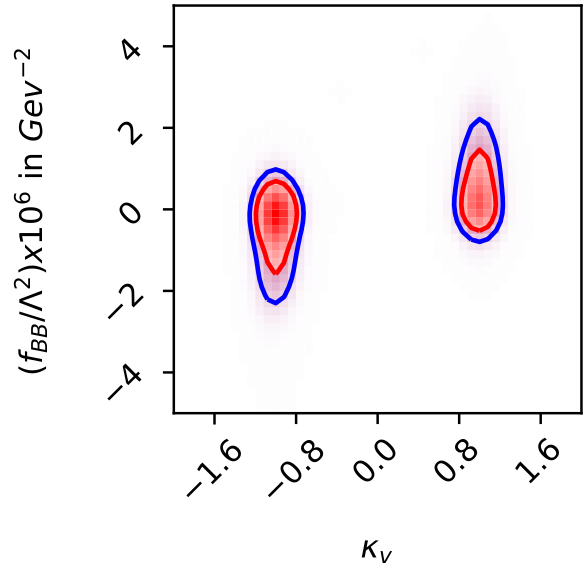}\hspace*{0.8cm}
\includegraphics[width=5.9cm, height=4.9cm]{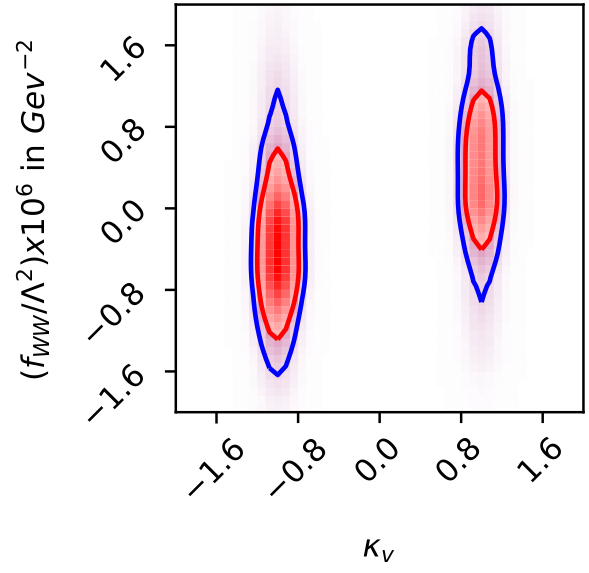}
\includegraphics[width=5.9cm, height=4.9cm]{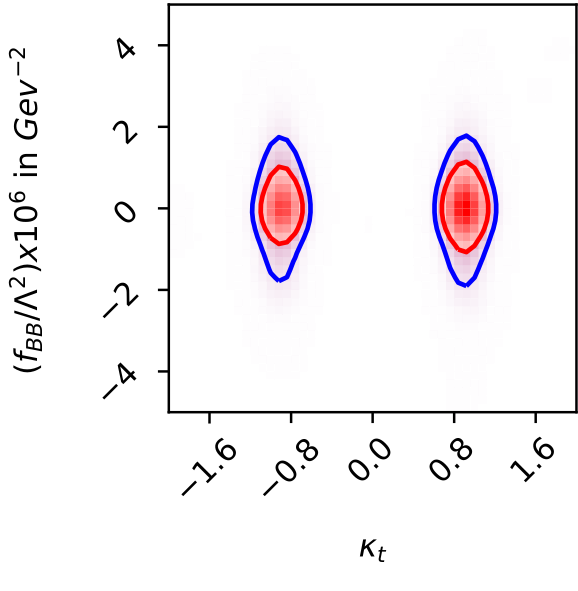}\hspace*{0.8cm}
\includegraphics[width=5.9cm, height=4.9cm]{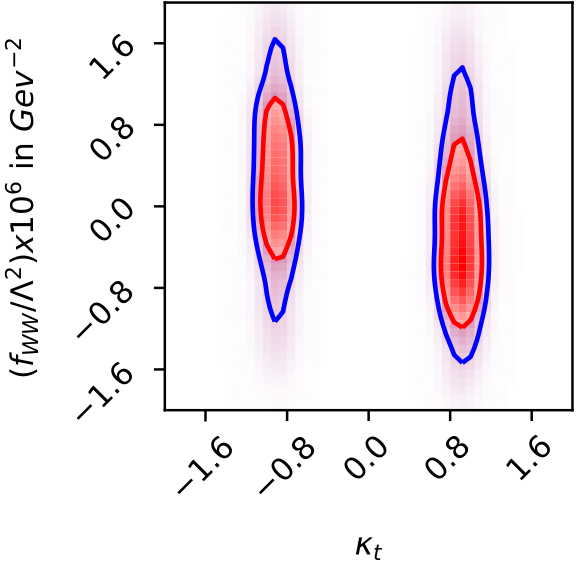}
\includegraphics[width=5.9cm, height=4.9cm]{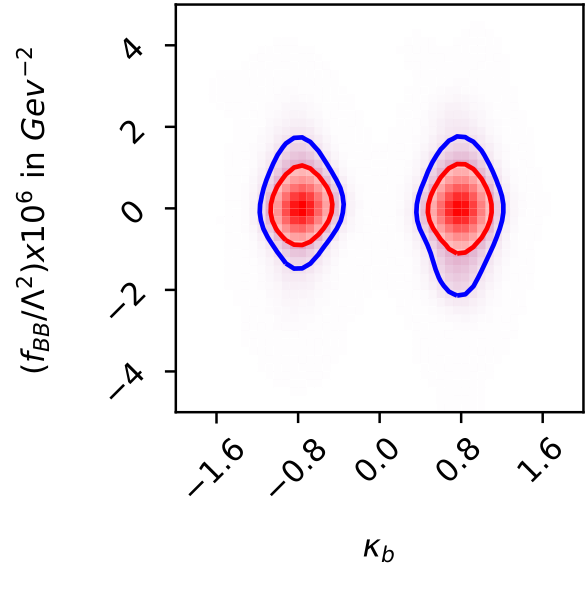}\hspace*{0.8cm}
\includegraphics[width=5.9cm, height=4.9cm]{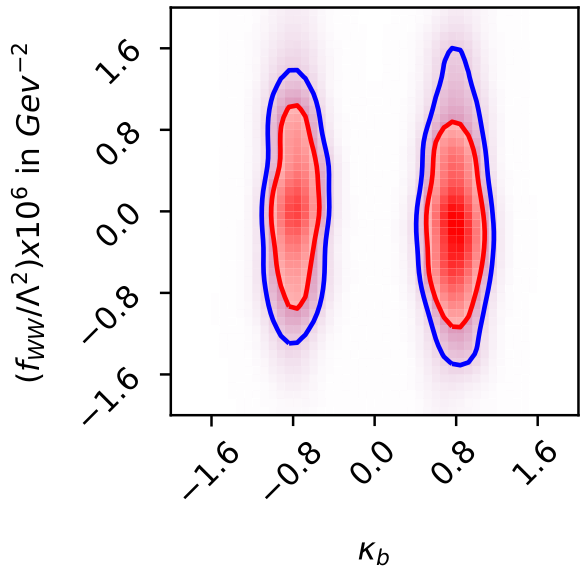}
\includegraphics[width=5.9cm, height=4.9cm]{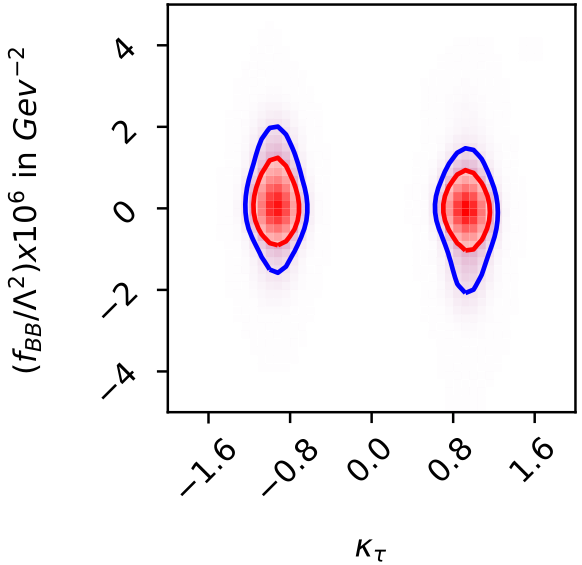}\hspace*{0.8cm}
\includegraphics[width=5.9cm, height=4.9cm]{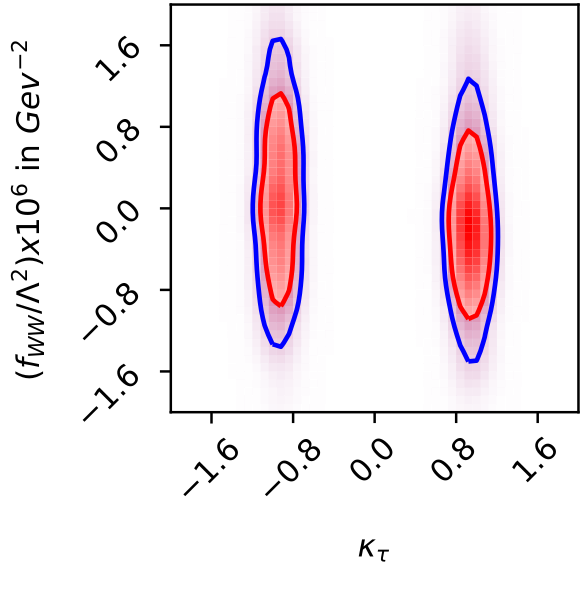}
\caption{Allowed regions at $1\sigma$(red) and $2\sigma$(blue) levels in the parameter space of scale factors and dimension-6 couplings}
\label{figure3}
\end{figure*}

\begin{table}
\centering
\begin{tabular}{| c | c | c | c |}
\hline \textbf{Channel} & \textbf{Experiment} & \textbf{Energy (GeV)}
& \textbf{Luminosity} \\ \hline $h \rightarrow \gamma \gamma$ & ATLAS
+ CMS~\cite{Khachatryan:2016vau} & 8 TeV & 19.5 fb$^{-1}$ \\ \hline $h
\rightarrow ZZ$ & ATLAS + CMS~\cite{Khachatryan:2016vau} & 8 TeV &
19.5 fb$^{-1}$ \\ \hline $h \rightarrow WW$ & ATLAS +
CMS~\cite{Khachatryan:2016vau} & 8 TeV & 19.5 fb$^{-1}$ \\ \hline $h
\rightarrow b \bar b$ & ATLAS + CMS~\cite{Khachatryan:2016vau} & 8 TeV
& 19.5 fb$^{-1}$ \\ \hline $h \rightarrow \tau \bar{\tau}$ & ATLAS +
CMS~\cite{Khachatryan:2016vau} & 8 TeV & 19.5 fb$^{-1}$ \\ \hline $h
\rightarrow \gamma \gamma$ & CMS~\cite{CMS:2017rli} & 13 TeV & 36
fb$^{-1}$ \\ \hline $h \rightarrow ZZ$ & CMS~\cite{CMS:2017jkd} &13
TeV & 36 fb$^{-1}$ \\ \hline $h \rightarrow WW$ &
CMS~\cite{CMS:2017pzi} & 13 TeV & 15.2 fb$^{-1}$ \\ \hline $h
\rightarrow \tau \bar{\tau}$ & CMS~\cite{Sirunyan:2017khh} & 13 TeV &
36 fb$^{-1}$ \\ \hline $h \rightarrow \gamma \gamma$ &
ATLAS~\cite{ATLAS-CONF-2017-045} & 13 TeV & 36 fb$^{-1}$ \\ \hline $h
\rightarrow ZZ$ & ATLAS~\cite{ATLAS-CONF-2017-043} &13 TeV & 36
fb$^{-1}$ \\ \hline
\end{tabular}
\caption{Listing of the Higgs signal strength measurements at LHC at 8
  and 13 TeV}
\label{tab1}
\end{table}

%We perform a detailed scan of our parameter space and calculate signal strength $\mu = \sigma_i \times BR_j$, where $i$ and $j$ indices denote the production and decay channel of Higgs boson respectively. Thereafter, taking into account of all the aforementioned experimental results(Table.~\ref{tab1}), we compute the minimum of the quantity $\chi^2$ and thereby obtain the best-fit value of the parameters. Then we identify the region which is allowed within 1 and 2$\sigma$ limit of the best-fit value. We have performed this analysis by the Monte-Carlo based $\chi^2$ minimization package MCMC~\cite{ForemanMackey:2012ig}. Here we present the plots showing the parameter space allowed at 1 and 2$\sigma$ by the experimental data. While plotting any allowed two-parameter space, we marginalize over all the remaining parameters.

%%%%%%%%%%%%%%%%%%%   ADD BEYOND THIS TO THE PRD VERSION %%%%%%%%%%%%%%%%%%%%

All 8 and 13 TeV results from ATLAS and CMS available so far are
included in our global fits. We thus take into account
the $h \rightarrow \gamma \gamma, ZZ, WW, b \bar b$ and $\tau \bar
\tau$ data from ATLAS and CMS at 8 TeV with 20fb$^{-1}$ integrated
luminosity and $h \rightarrow \gamma \gamma, ZZ, WW, \tau \bar{\tau}$
from CMS 13 TeV data with 36fb$^{-1}$ integrated luminosity and $h
\rightarrow \gamma \gamma, ZZ$ from ATLAS 13 TeV data with 36 fb$^{-1}$ integrated
luminosity. These details are presented in Table.~\ref{tab1}.

We  calculate the  signal
strengths $\mu = \sigma_i \times BR_j$, where the indices $i$ and $j$ indices
denote respectively the production and decay channels of Higgs boson.
One then proceeds to obtain $\chi^2_{min}$, based on the 
experimental best-fit values and the corresponding
error intervals, whence one arrives at plots showing
the hypersurfaces allowed at the 1-and 2$\sigma$ levels. 
Projections of these hypersurfaces into various two-parameter 
subspaces are shown in Figures 2, 3 and 4, where 
all remaining parameters have been marginalized.

 Figure~\ref{figure2} includes 2$\sigma$ contours in various pairs of the
 scaling parameters $\kappa$.  The quadratic nature of the
 dependence of $\chi^2$ on the $\kappa$'s results in the
 symmetric nature of the contours.

In Figure~\ref{figure3} we show correlation between the $\kappa$'s and the
Wilson coefficients $f_{BB}$ and $f_{WW}$.  $f_B$ and $f_W$ contribute
to $h \rightarrow Z \gamma$ but not $h \rightarrow \gamma \gamma$,
since only the neutral component of the Higgs doublet acquires a vev.
Therefore, they are  {\em not} 
constrained by the observed signal strengths, but have  only
upper limits from the non-observation of the $Z\gamma$ final state so far.

\begin{figure*}[!hptb]
\centering
\includegraphics[width=5.9cm, height=4.9cm]{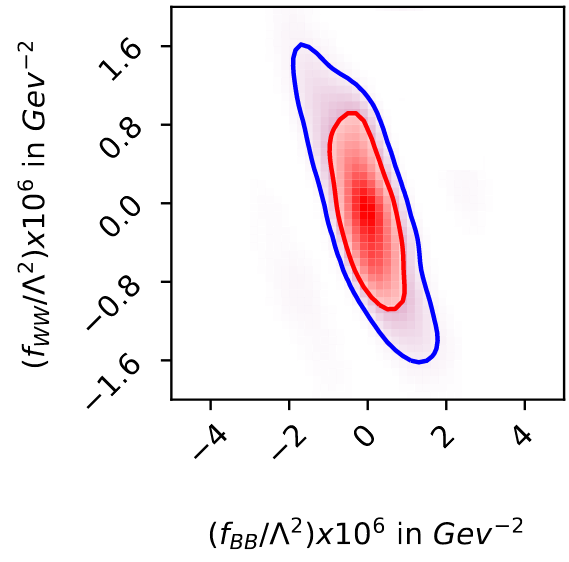}\hspace*{0.8cm}
\includegraphics[width=5.9cm, height=4.9cm]{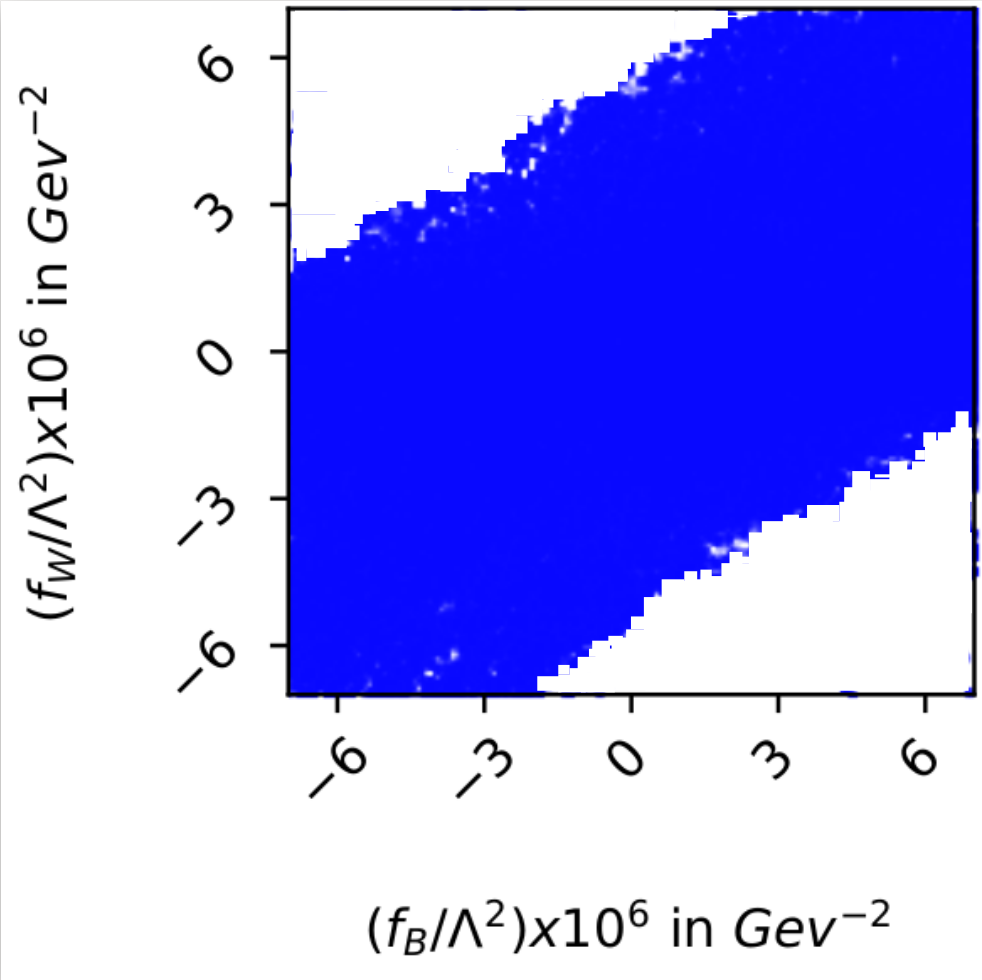}
\caption{(left) Allowed regions at $1\sigma$(red) and $2\sigma$(blue) levels in the $f_{BB}$-$f_{WW}$ plane and (right) allowed region at 2$\sigma$ level in the $f_B$-$f_W$ plane.}
\label{figure4}
\end{figure*}

Figure~\ref{figure4} (left) shows the $2\sigma$ region in the
$f_{BB}$-$f_{WW}$ plane.  There is a direction in this plane, where
the signal strength in the $h \rightarrow \gamma \gamma$ channel will
always be satisfied, as the contribution to the diphoton final state
depends on $f_{BB}+f_{WW}$. This is evident from
Table.~\ref{obbww}. On the other hand, the process $h \rightarrow Z
\gamma$ receives contribution via the combination $\tan^2 \theta_W
f_{BB}-f_{WW}$, thus indicating a region in the same place when the
latter process fails to be restrictive. It is only from a combined
imposition of both constraints that one obtains a closed region in the
$f_{BB}$-$f_{WW}$, as can be seen in Figure~\ref{figure4} (left).
Figure~\ref{figure4} (right) shows the favored regions in the
$f_B$-$f_W$ plane. As $f_B$ and $f_W$ get contribution only from the
upper limit of $h \rightarrow Z \gamma$ process~\cite{Aaboud:2017uhw}, we do not get a
closed region in this case.

\section{Extended 
Higgs models and dimension-6 operators}\label{model}

Various new physics models
predict extended electroweak symmetry breaking sectors. 
It is naturally of interest to link the model-independent analysis 
presented above to  specific theoretical scenarios. It has to be 
remembered that most of these scenarios are strongly constrained by not only
electroweak precision data but also observed signal strengths of the
already discovered Higgs. From the standpoint of uncovering new theoretical
structures, it is therefore a challenge to extract the tiniest tractable
departures, in multiplicative alteration of interaction strength(s) 
and, perhaps more significantly, in traces of higher dimensional
couplings. With this in view, we now translate the results of the previous
section to those pertaining to extended Higgs models, taking into account
the additional constraints that connect model parameters in each case.

The simplest set of extensions consists in two Higgs doublet models (2HDM). 
Motivated and largely popularized by the
minimal supersymmetric standard Model (MSSM), such scenarios 
are of interest independently of supersymmetry. They are of interest due to  
their  phenomenological richness but also, for example, by the observation
that 2HDM allows a stable electroweak vacuum all the way to the Planck scale
without the aid of any new physics~\cite{Chakrabarty:2014aya}.

At the same time, scalars in higher $SU(2)$ representations
are often helpful in understanding basic issues such as neutrino mass
generation. Introduction of a $Y=2$ complex $SU(2)$ triplet enables the 
Type II seesaw mechanism to yield Majorana masses for neutrinos.
The strong constraint on the triplet vev  from the $\rho$-parameter 
is not only consistent with such mechanism but in fact justifies
the smallness of neutrino masses. Furthermore, some model-building
efforts to connect neutrino masses with their mixing angles motivates
Type II seesaw with {\em two triplets instead of one}.

\begin{figure}[!hptb]
\centering
\hspace{-2cm}
\includegraphics[width=12.8cm, height=9.3cm]{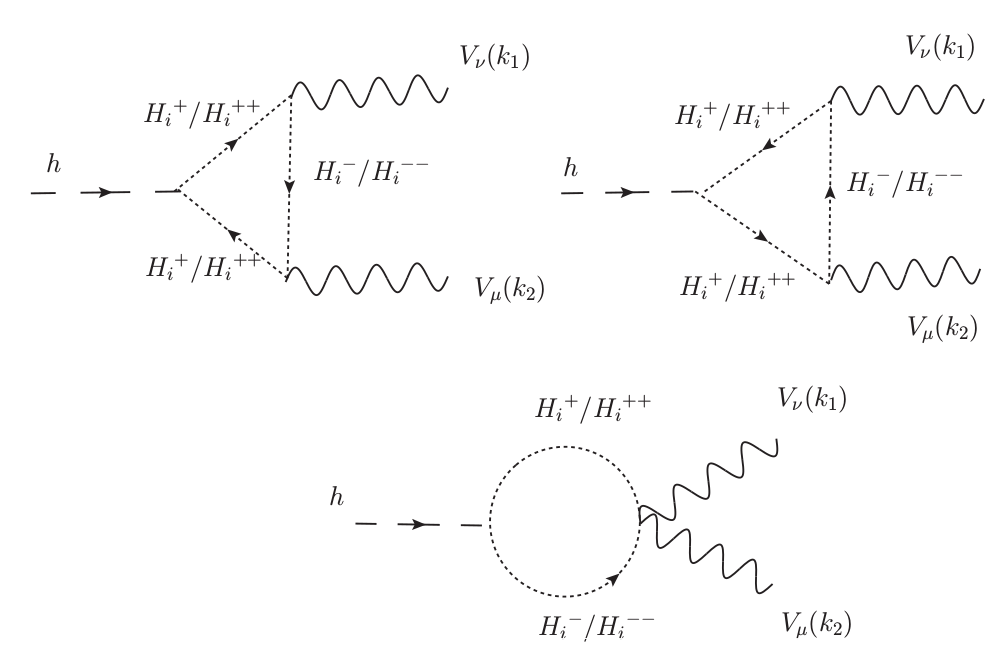}
\caption{Additional scalar loop contributions to $h \rightarrow \gamma \gamma$ 
and $h \rightarrow Z \gamma$.}
\label{figure5}
\end{figure}

We analyze some of these models here, with reference to the
model-independent approach of the previous section. 
All scenarios discussed here predict one or
more, singly or doubly, charged Higgs bosons. These
charged scalars  should contribute to  loop-induced decays of
the SM-like Higgs,  namely, $h \rightarrow \gamma \gamma$, $h \rightarrow Z
\gamma$. The corresponding decay
widths deviate from the SM predictions due to (a) scaling of
the $hWW, ht\bar{t}$ vertices, and (b) the additional contributions
from diagrams of the kind shown in Figure~\ref{figure5}, where the $V^{\mu}$ and $V^{\nu}$ in the external legs stand for $\gamma \gamma$ or $\gamma Z$. While the former
lead to the factors denoted by $\kappa$, the latter are parametrized in
our approach by dimension-6 operators. The difference compared to the
model-independent case is that the two classes of modifiers are related
once model parameters are specified. Such mutual dependence has to be taken
into account in each particular situation.

The contribution to the amplitude from the charged scalar loops 
for  $h \rightarrow \gamma \gamma$ is of the form
\begin{equation}
-i{\cal M} = C_{vertex}\times(k_1.k_2 g_{\mu \nu} -
k_{1\mu}k_{2\nu})\epsilon^{*\mu}(k_2)\epsilon^{*\nu}(k_1)\times
A_H(\gamma \gamma)(m_{H^{\pm}})
\label{gammamod}
\end{equation}
While that for 
$h \rightarrow Z \gamma$ is
\begin{equation}
-i{\cal M} = \tilde{C}_{vertex}\times(k_1.k_2 g_{\mu \nu} -
k_{1\mu}k_{2\nu})\epsilon^{*\mu}(k_2)\epsilon^{*\nu}(k_1)\times A_H(Z
\gamma)(m_{H^{\pm}},m_Z)
\label{zgammod}
\end{equation}

Where $C_{vertex}$ and $\tilde{C}_{vertex}$ are the vertex factors for
$h \rightarrow \gamma \gamma$ and $h \rightarrow Z \gamma$
in the two cases, while $A_H(\gamma \gamma$) and $A_H(Z
\gamma$) are the loop integrals for $h \rightarrow \gamma \gamma$ and
$h \rightarrow Z \gamma$ respectively.

The current direct search and B-physics observables do not put strong
limits on charged Higgs masses except in very specific models.
Therefore, the charged scalars need not always be much heavier than
the electroweak scale, and can easily be just a few hundred GeV in
mass. 

Therefore, while parametrizing the amplitudes in terms of gauge
invariant effective operators, one may still have `effective'
suppression scales $\gtrsim$ TeV. The combination of Wilson coefficients
contributing to the decay amplitudes, evaluated at the EWSB scale, may in principle have a functional
dependence on $m_{EWSB} \simeq m_Z$ due to renormalization group (RG)
running. However, this dependence is rather weak in general, unless
one has strongly coupled high-scale dynamics.  

\begin{figure}[!hptb]
%\hspace{-2cm}
\centering
\includegraphics[width=11.5cm, height=6.0cm]{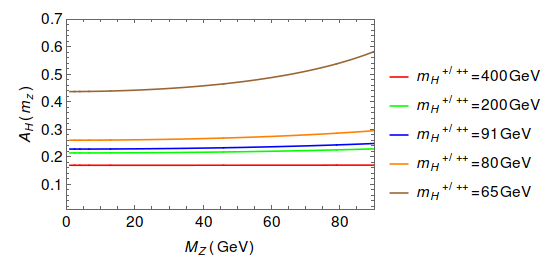}
\caption{Dependence of the additional scalar loop integral on the $Z$ boson mass in $h \rightarrow Z \gamma$.}
\label{figure6}
\end{figure}

In the cases under consideration here, the loop
amplitude for $h \rightarrow \gamma \gamma$ does not involve $m_Z$,
and the charged scalar masses arises from $SU(2)_L \times U(1)_Y$
invariant terms. Thus there is no dependence on $m_{EWSB}$ there. One
would thus expect the same dependence (or lack of it) in the amplitude
for $h \rightarrow Z \gamma$ unless there are highly fine-tuned
boundary conditions in the RG running of parameters. A natural way of
establishing consistency between the two amplitudes, therefore, is to have 
no $m_Z$-dependence in the loop amplitude for $h \rightarrow Z \gamma$ as
well.  As one can see from Figure~\ref{figure6}, this is indeed the
case here; the loop integrals are insensitive to an `artificial'
variation of $m_Z$ so long as the charged Higgs(es) circulating in the
loop stay above that mass.  It is, therefore, legitimate to
encapsulate the contributions to these loop amplitudes as Wilson
coefficients of dimension-6 gauge-invariant operators, so long as the
charged scalars responsible for these amplitude are at least a factor of
two heavier than the $Z$.

For reasons already mentioned, we adopt four kinds of 2HDM and also
scenarios with one and two scalar triplets to illustrate our approach.
We take them up in turn in the following subsections.
%{\bf Add some discussions of why $f_B , f_W$ have been fixed at their 
%best-fit values in the entire analysis following this.}
We mention here that in the analysis that follows, we have marginalized over the parameters $f_B$ and $f_W$. These two parameters get constrained only from the upper limit on $h \rightarrow Z \gamma$ decay width. Thus their role in this analysis is to stay within the boundary defined by this upper limit; no further information can be obtained at present due to absence of any positive result. 

\subsection{Two Higgs doublet models}

The most general 2HDM potential contains fourteen parameters~\cite{Branco:2011iw,Gunion:1989we}. It can have
both CP-conserving and CP-violating terms in principle. However, we
keep the analysis simple by assuming no CP-violation in the Higgs sector. 
The most general CP-conserving potential with two Higgs doublets
$\Phi_1$ and $\Phi_2$ is 

\begin{eqnarray}
V = m_{11}^2\Phi^{\dagger}_1 \Phi_1 + m_{22}^2\Phi^{\dagger}_2 \Phi_2
- m_{12}^2 \left(\Phi^{\dagger}_1 \Phi_2 + \Phi^{\dagger}_2 \Phi_1
\right) + \frac{\lambda_1}{2}\left(\Phi^{\dagger}_1 \Phi_1 \right)^2 +
\frac{\lambda_2}{2}\left(\Phi^{\dagger}_2 \Phi_2 \right)^2 \nonumber
\\ + \lambda_3 \Phi^{\dagger}_1 \Phi_1 \Phi^{\dagger}_2 \Phi_2 +
\lambda_4 \Phi^{\dagger}_1 \Phi_2 \Phi^{\dagger}_2 \Phi_1 +
\frac{\lambda_5}{2} \left[ \left(\Phi^{\dagger}_1 \Phi_2 \right)^2 +
  \left( \Phi^{\dagger}_2 \Phi_1 \right)^2 \right]
\label{higgspot}
\end{eqnarray}

where all coefficients are real. The minimum
of this potential is obtained when the neutral 
components of $\Phi_1$ and $\Phi_2$ acquire vev $v_1$ 
and $v_2$ respectively.

\beq
{\braket{\phi_1}}_0 = \frac{1}{\sqrt 2}\left( 
\begin{array}{c}
0 \\ v_1 \\  
\end{array} \right)
 \eeq,\beq
{\braket{\phi_2}}_0 = \frac{1}{\sqrt 2}\left( 
\begin{array}{c}
0 \\ v_2 \\  
\end{array} \right)
 \eeq

\beq
\phi_a(x)= \frac{1}{\sqrt 2}\left( 
\begin{array}{c}
\phi_a^+ (x) \\ v_a + \sigma_a (x) + i \eta_a (x) \\  
\end{array} \right)
 \eeq Where a=1,2. After substituting this form of $\Phi_{1,2}$ in
 Equation~\ref{higgspot}, we obtain the mass matrices for the charged and
 neutral Higgs bosons. Their diagonalization yields the masses of
the charged scalar, two neutral scalars as well as the neutral pseudoscalar.
The charged scalar mass in particular is given by

\beq
m_{H^{\pm}}^2 = [m_{12}^2/(v_1v_2) - \lambda_4 - \lambda_5](v_1^2+v_2^2)
\eeq
These masses, together with the neutral scalar mixing angle $\alpha$ and
charged and pseudoscalar mixing angle $\beta$ ($\tan \beta = \frac{v_2}{v_1}$) constitute the set of free parameters in
this model.

The Yukawa sector is what mainly distinguishes one subclass
of 2HDM from another. These subclasses have developed largely 
from different discrete symmetries postulated to address the issue 
of avoiding tree-level flavor changing neutral currents (FCNC). 
The most general  Yukawa Lagrangian in a 2HDM  can be expressed as

\begin{equation} 
\label{Yukawa}
{\cal L} = y^1_{ij}\bar Q_{iL} \Phi_1 d_{jR} + y^2_{ij}\bar Q_{iL}
{\bar{\Phi}}_1 u_{jR} + y^3_{ij}\bar Q_{iL} \Phi_2 d_{jR} +
y^4_{ij}\bar Q_{iL} {\bar{\Phi}}_2 u_{jR} + y^5_{ij}\bar L_{iL} \Phi_1
e_{jR} + y^6_{ij}\bar L_{iL} \Phi_2 e_{jR}
\end{equation}

FCNC constraints discourage the coupling of both doublets with $T_3 =
+$ and$-$ fermions. The most popular way to circumvent this problem
is to impose a $Z_2$ symmetry, thereby prohibiting some of the terms
above. Different combinations of the $Z_2$ quantum numbers of the
fields present in Equation~(\ref{Yukawa}) lead to at least four types of 2HDM,
which are brought under our analysis in turn.

In the subsequent discussions we will mainly focus on Higgs decay to $\gamma
\gamma$ and $Z \gamma$ channels, which is dominated by top-and $W$-loops 
in SM. In all the variants of 2HDM,
the charged Higgs boson ($H^{\pm}$) contributes additionally to this
processes at the one-loop level. The modification of the $W$ and top loop over and above their SM values and the additional charged Higgs loop contribution is given in detail in Appendix~\ref{app1}. 

\subsubsection{Type-I 2HDM}

Here all fermions are assumed to couple to the same doublet, so that Equation~(\ref{Yukawa}) reduces to

\begin{equation}
\label{type1}
{\cal L}_{Yukawa} = y^1_{ij}\bar Q_{iL} \Phi_2 d_{jR} + y^2_{ij}\bar
Q_{iL} {\bar{\Phi}}_2 u_{jR} + y^5_{ij}\bar L_{iL} \Phi_2 e_{jR}
\end{equation}
This can be achieved by imposing the discrete symmetry on the 
${\cal L}_{Yukawa}$, $\Phi_1 \rightarrow -\Phi_1$.  
The couplings of the fermions with the SM-like Higgs boson in this
case are

\begin{eqnarray}
C_{h t \bar t} = \frac{\cos \alpha}{\sin \beta} \times C_{h t \bar
  t}^{\text SM} \nonumber \\ C_{h b \bar b} = \frac{\cos \alpha}{\sin
  \beta} \times C_{h b \bar b}^{\text SM} \nonumber \\ C_{h \tau \bar
  \tau} = \frac{\cos \alpha}{\sin \beta} \times C_{h \tau \bar
  \tau}^{\text SM}
\label{modtype1}
\end{eqnarray}
while the gauge boson couplings are

\beq
C_{hVV} = \sin(\beta - \alpha) \times C_{hVV}^{\text SM} 
\eeq

\begin{figure}[!htb]
\begin{minipage}[!htb]{0.45\linewidth}
\centering
%\vspace*{0.7cm}
%\psfrag{ml}{$\bf{m_{\tilde{e}_L}} $}
\includegraphics[width=6.5cm, height=5.15cm]{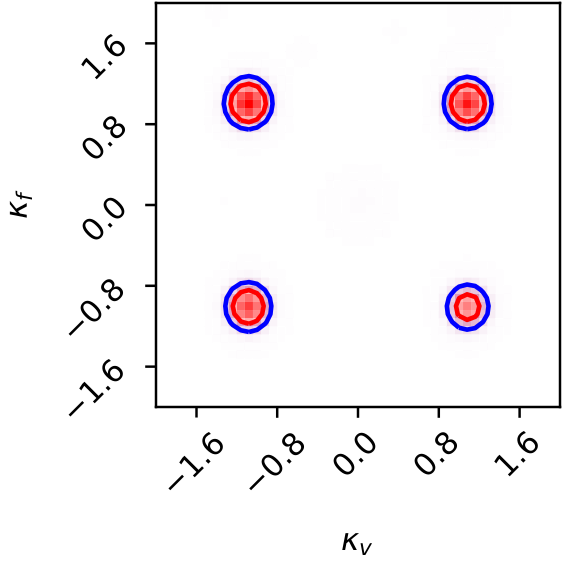}
%\caption{}
%\label{fig:mssm_ln}
\end{minipage}
\hspace{0.9cm}
\begin{minipage}[!htb]{0.45\linewidth}
\centering
%\psfrag{mr}{$\bf{m_{\tilde{e}_R}} $}
\includegraphics[width=6.5cm, height=5.15cm]{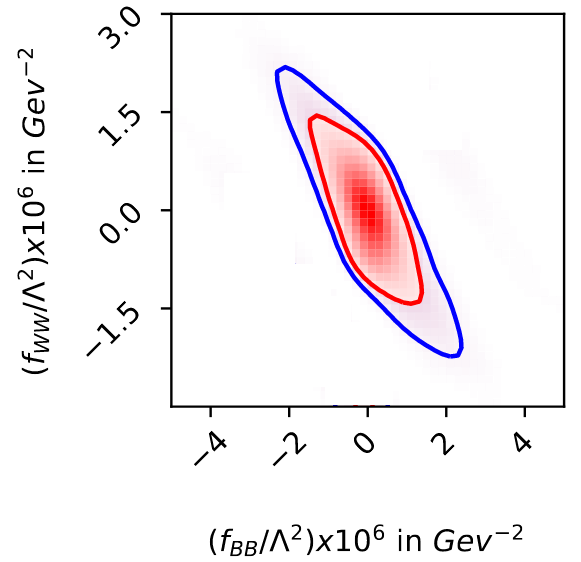}
%\caption{}
%\label{fig:mssm_br}
\end{minipage}
\caption{Allowed regions at $1\sigma$(red) and $2\sigma$(blue)
  levels in the parameter space of scale factors and dimension-6
  couplings in Type I 2HDM}
\label{figure7}
\end{figure}

These define, in this case as well as the other types of 2HDM, the
corresponding $\kappa_v$, $\kappa_t$, $\kappa_b$ and $\kappa_{\tau}$.
Here they all of the same value, denoted by the factor $\kappa_f$.
Thus the departure from SM effects in data concerning the 125 GeV scalar  
can be described here in terms of six parameters, namely,
$\kappa_v, \kappa_f, f_{BB}, f_{WW}, f_B, f_W$. Following the same
procedure as in Section~\ref{np} and assuming all the six parameters
to be independent, the $1\sigma$ and 
$2\sigma$  allowed regions for them are found to be as in
Figure~\ref{figure7}, which looks very similar to the corresponding
contours in Figures~\ref{figure2} and \ref{figure4}.

\begin{figure}[!htb]
\begin{minipage}[!htb]{0.45\linewidth}
\centering
%\vspace*{0.7cm}
%\psfrag{ml}{$\bf{m_{\tilde{e}_L}} $}
\includegraphics[width=6.5cm, height=5.15cm]{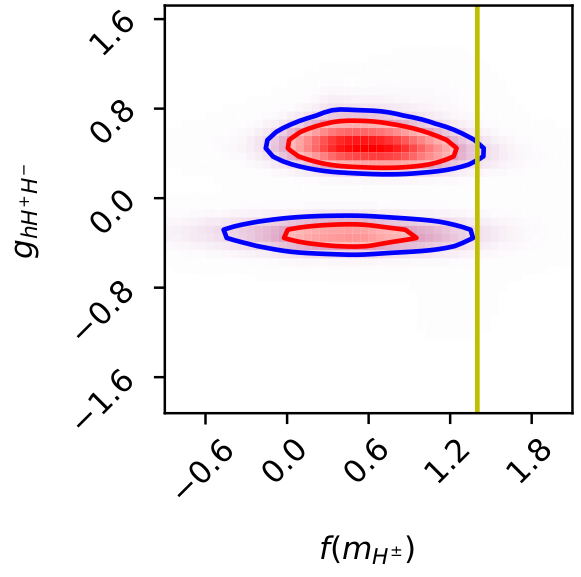}
%\caption{}
%\label{fig:mssm_ln}
\end{minipage}
\hspace{0.9cm}
\begin{minipage}[!htb]{0.45\linewidth}
\centering
%\psfrag{mr}{$\bf{m_{\tilde{e}_R}} $}
\includegraphics[width=6.5cm, height=5.15cm]{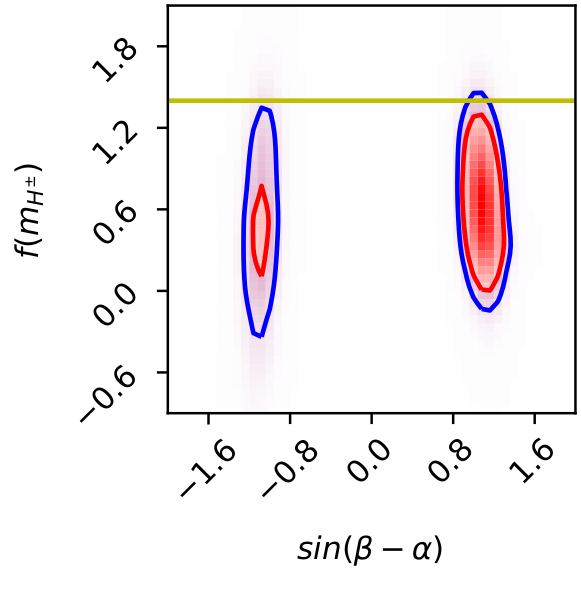}
%\caption{}
%\label{fig:mssm_br}
\end{minipage}
\caption{Allowed regions at $1\sigma$(red) and $2\sigma$(blue)
  levels in the parameter space of $f(m_{H^{\pm}})-g_{hH^+H^-}$ (left) and  $\sin(\beta-\alpha)-f(m_{H^{\pm}})$ (right) in Type I 2HDM}
\label{figure8}
\end{figure}   

The task now is to translate these into  limits on the model parameter space. 
In order to do that, we note that in this model,

\begin{eqnarray}
\kappa_v &=& \sin(\beta - \alpha) \\ \kappa_f &=& \frac{\sin
  \alpha}{\cos \beta} = \sin(\beta - \alpha) - \cos(\beta - \alpha)
\tan \beta
\end{eqnarray}
 The top, bottom and $W$ loop contribution to $h \rightarrow \gamma
 \gamma$ will be modified by these scale factors. 
 The charged Higgs loop contribution will be a
 function of $m_{H^{\pm}}$ multiplied by the quantity $g_{hH^+H-}$ which 
 is given by

\beq g_{hH^+H^-} = \sin(\beta - \alpha) + \frac{\cos2\beta \sin(\beta+
  \alpha)}{2\cos^2\theta_W} \eeq

At this point we draw the attention of the reader to Figure~\ref{figure7}.
Unlike in the model-independent case, we find this time an interrelation
of $f_{BB}, f_{WW}$ and $\kappa$, since $\alpha$ and $\beta$ enter into all 
of them. At the same time, there is a single $\kappa$ here, pointing
again at a correlation among the scaling factors. 

The fact that the top-quark coupling plays the most crucial role
here in fitting the data shows little departure in the left panel of 
Figure~\ref{figure7}, as compared to the corresponding contour in Figure~\ref{figure2}.
As for the right panel, while $f_{BB}, f_{WW}$ is correlated with
$\kappa$, the availability of an extra parameter
$m_{H^{\pm}}$ softens this correlation, since the entire $2\sigma$ contour 
in $f_{BB}-f_{WW}$ plane will be allowed for some value
of $m_{H^{\pm}}$. That is why the right panel of  Figure~\ref{figure7}
is very similar to the corresponding figure in Figure~\ref{figure4}, modulo
the fact that here is only $\kappa_v$ and $\kappa_f$ that are marginalized. 

For the sake
of simplicity in the analytical calculation we define a parameter
which is a function of the charged Higgs mass:

\begin{equation}
f(m_{H^{\pm}}) = \tau f^0(\tau)
\label{fmHp}
\end{equation}

%{\bf Make $f' --> f^0 $everywhere}

where $\tau$ and $f^0(\tau)$ have been defined in Equations~\ref{tau} and
~\ref{ftau}. This function encapsulates the contribution
of the charged Higgs loop integral.

Keeping the above discussion in mind, one arrives at the $1\sigma$ and
$2\sigma$ contours in the $\sin(\beta-\alpha)$-$f(m_{H^{\pm}})$ and
$g_{hH^+H^-}$-$f(m_{H^{\pm}})$ planes in Figure~\ref{figure8} (left) and
(right) respectively. We mention here in the context of loop induced
effects, $f(m_{H^{\pm}})$ can only take positive values, as otherwise
one ends up with an inadmissibly small charged Higgs mass (see
Equation~\ref{ftau}).

%............................

\subsubsection{Type-II 2HDM}

Another way to stay away from tree-level FCNC is to allow up-type
quarks to couple to one doublet, and down-type quarks and leptons to
couple to another. Under this assumption, the ${\cal L}_{Yukawa}$
becomes

\begin{equation}
\label{type2}
{\cal L}_{Yukawa} = y^1_{ij}\bar Q_{iL} \Phi_1 d_{jR} + y^2_{ij}\bar
Q_{iL} {\bar{\Phi}}_2 u_{jR} + y^5_{ij}\bar L_{iL} \Phi_1 e_{jR}
\end{equation}
 This can be enforced by demanding that the ${\cal L}_{Yukawa}$
 remains invariant under $\Phi_1 \rightarrow -\Phi_1$ and $d_R
 \rightarrow -d_R$ and $e_R \rightarrow -e_R$.  

 The SM-like Higgs boson now has scaled interactions with SM fermions
 and gauge bosons given by

\begin{eqnarray}
C_{h t \bar t} = \frac{\cos \alpha}{\sin \beta} \times C_{h t \bar
  t}^{\text SM} \nonumber \\ C_{h b \bar b} = -\frac{\sin \alpha}{\cos
  \beta} \times C_{h b \bar b}^{\text SM} \nonumber \\ C_{h \tau \bar
  \tau} = -\frac{\sin \alpha}{\cos \beta} \times C_{h \tau \bar
  \tau}^{\text SM} \nonumber \\ C_{hVV} = \sin(\beta - \alpha) \times
C_{hVV}^{\text SM}
\label{mod2}
\end{eqnarray}

The top, bottom and $\tau$ couplings with Higgs are not
modified in the same way any more. However, Equation~\ref{mod2}.
implies 

\begin{eqnarray}
\kappa_{\tau} = \kappa_b = \frac{\kappa_t^2 - 2 \kappa_t \kappa_v
  +1}{\kappa_t - \kappa_v} 
\label{ktau2}
\end{eqnarray}

\begin{figure}[!htb]
\begin{minipage}[!htb]{0.45\linewidth}
\centering
%\vspace*{0.7cm}
%\psfrag{ml}{$\bf{m_{\tilde{e}_L}} $}
\includegraphics[width=6.5cm, height=5.15cm]{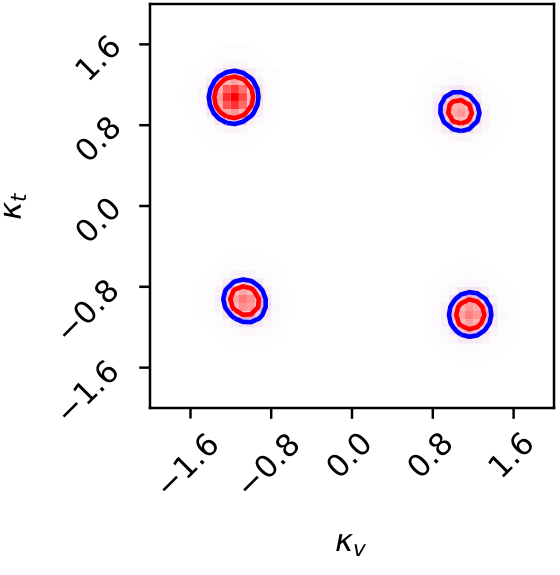}
%\caption{}
%\label{fig:mssm_ln}
\end{minipage}
\hspace{0.9cm}
\begin{minipage}[!htb]{0.45\linewidth}
\centering
%\psfrag{mr}{$\bf{m_{\tilde{e}_R}} $}
\includegraphics[width=6.5cm, height=5.15cm]{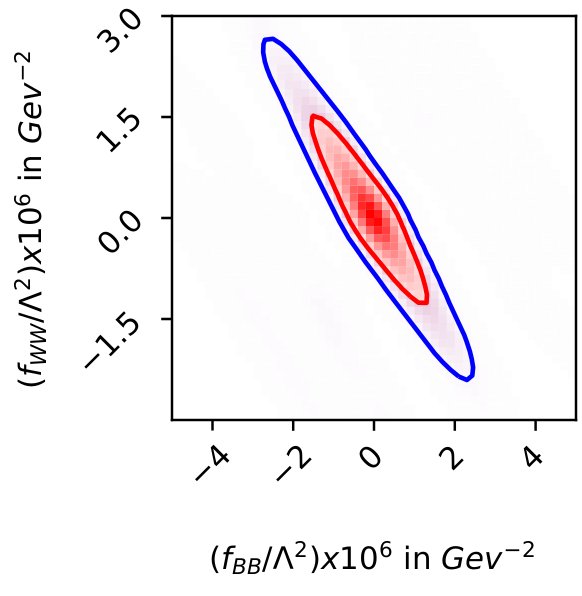}
%\caption{}
%\label{fig:mssm_br}
\end{minipage}
\caption{Allowed regions at $1\sigma$(red) and $2\sigma$(blue)
  levels in the parameter space of scale factors and dimension-6
  couplings in Type II 2HDM}
\label{figure9}
\end{figure}

On using these relations, we can again reduce the 8-dimensional
parameter space to a 6-dimensional one, albeit via different constraints. 
The independent parameters are once more
$\kappa_v, \kappa_t, f_{BB}, f_{WW}, f_B$ and $f_W$. The
allowed parameter space at $1\sigma$ and $2\sigma$ regions. In $\kappa_v$- $\kappa_t$ and $f_{BB}$-$f_{WW}$ parameter space can be seen
in Figure~\ref{figure9} (left) and (right) respectively.

One now translates these limits into those on the model parameter space,
using

\begin{eqnarray}
\kappa_v &=& \sin(\beta - \alpha) \\ \kappa_t &=& \frac{\cos
  \alpha}{\sin \beta} = \sin(\beta - \alpha) - \cos(\beta - \alpha)
\tan \beta
\end{eqnarray}

\begin{figure}[!htb]
\begin{minipage}[!htb]{0.45\linewidth}
\centering
%\vspace*{0.7cm}
%\psfrag{ml}{$\bf{m_{\tilde{e}_L}} $}
\includegraphics[width=6.5cm, height=5.15cm]{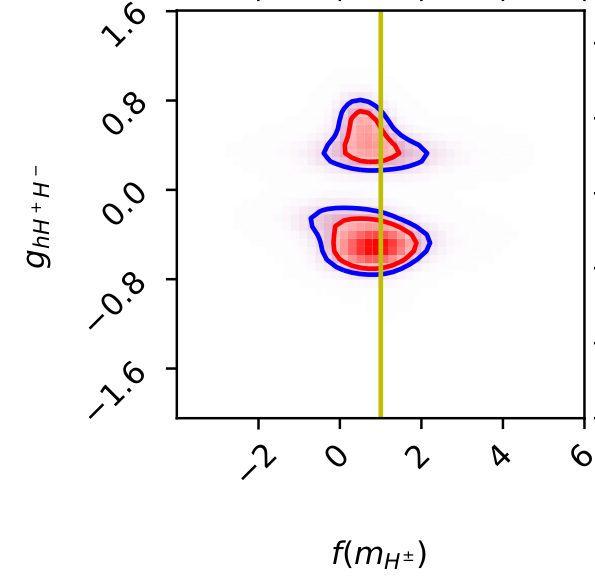}
%\caption{}
%\label{fig:mssm_ln}
\end{minipage}
\hspace{0.9cm}
\begin{minipage}[!htb]{0.45\linewidth}
\centering
%\psfrag{mr}{$\bf{m_{\tilde{e}_R}} $}
\includegraphics[width=6.5cm, height=5.15cm]{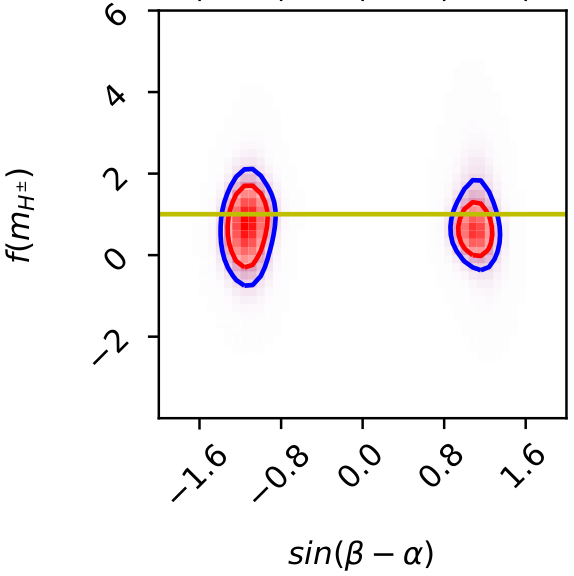}
%\caption{}
%\label{fig:mssm_br}
\end{minipage}
\caption{Allowed regions at $1\sigma$(red) and $2\sigma$(blue)
  levels in the parameter space of $f(m_{H^{\pm}})-g_{hH^+H^-}$ (left) and  $\sin(\beta-\alpha)-f(m_{H^{\pm}})$ (right) in Type II 2HDM}
\label{figure10}
\end{figure}

$\kappa_b$ and $\kappa_{\tau}$ can be obtained following 
Equations~\ref{mod2} and~\ref{ktau2}. The top, bottom and $W$ loop
contribution to the decay $h \rightarrow \gamma \gamma$ will be
modified by scale factors as described above. The new charged
Higgs loop contribution, too, will depend on $\sin(\beta - \alpha)$,
$\tan \beta$ and $m_{H^{\pm}}$ as discussed in the previous
subsection. A similar procedure as that for Type-I 2HDM
leads to the contours in Figure~\ref{figure10}.

It should be noted that contours in the $\kappa_v-\kappa_t$ here   
exhibit more departure from the corresponding figure in Figure~\ref{figure2}
than is seen for Type-I 2HDM. This is because the correlated
scale factors $\kappa_b$ and $\kappa_{\tau}$ have a stronger role 
to play here, as they make significant contributions to loop-induced
Higgs decays for high values of $\tan\beta$, thus causing shifts in
the allowed contours. In addition, the allowed region in the right panel
of Figure~\ref{figure10} suffers from stronger constraints than in the
previous case, as the charged Higgs mass has a lower limit of
580 GeV in Type-II 2HDM from the rare decay $b\rightarrow s\gamma$~\cite{Haisch:2008ar,Mahmoudi:2009zx,Gupta:2009wn,Jung:2010ik,Misiak:2015xwa}.

\subsubsection{Lepton-specific 2HDM}

An alternative way is to allow the up-and
down-type quarks to couple to one doublet and leptons to the other doublet. 
This can be achieved by imposing the symmetry of the Lagrangian under the
discrete symmetry $\Phi_1 \rightarrow -\Phi_1$ and $e_R \rightarrow
-e_R$. ${\cal L}_{Yukawa}$ in this case becomes

\begin{equation}
\label{lepton}
{\cal L}_{Yukawa} = y^1_{ij}\bar Q_{iL} \Phi_2 d_{jR} + y^2_{ij}\bar
Q_{iL} {\bar{\Phi}}_2 u_{jR} + y^5_{ij}\bar L_{iL} \Phi_1 e_{jR}
\end{equation}

In this scenario the scale factors which modify the SM Higgs
coupling with fermions and gauge bosons, take the following forms:

\begin{eqnarray}
C_{h t \bar t} = \frac{\cos \alpha}{\sin \beta} \times C_{h t \bar t}^{\text SM}  \nonumber \\
C_{h b \bar b} = \frac{\cos \alpha}{\sin \beta} \times C_{h b \bar b}^{\text SM}   \nonumber \\
C_{h \tau \bar \tau} = -\frac{\sin \alpha}{\cos \beta} \times C_{h \tau \bar \tau}^{\text SM}   \nonumber \\
C_{hVV} = \sin(\beta - \alpha) \times C_{hVV}^{\text SM} 
\label{mod3}
\end{eqnarray}

The up and down-type quark couplings thus get modified in
the same manner, whereas the lepton coupling is different. Going back
to our parametrization of new physics in the model-independent framework,

\begin{eqnarray}
\kappa_t &=& \kappa_b  \\
\kappa_{\tau} &=& \frac{\kappa_t^2 - 2 \kappa_t \kappa_v +1}{\kappa_t - \kappa_v}
\end{eqnarray}

\begin{figure}[!htb]
\begin{minipage}[!htb]{0.45\linewidth}
\centering
%\vspace*{0.7cm}
%\psfrag{ml}{$\bf{m_{\tilde{e}_L}} $}
\includegraphics[width=6.5cm, height=5.15cm]{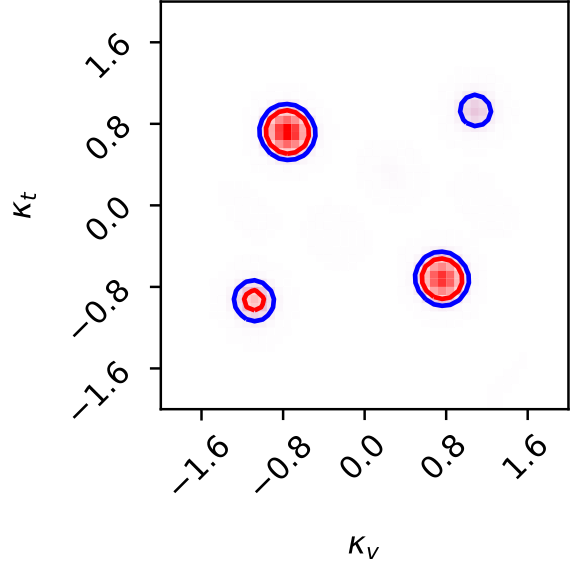}
%\caption{}
%\label{fig:mssm_ln}
\end{minipage}
\hspace{0.9cm}
\begin{minipage}[!htb]{0.45\linewidth}
\centering
%\psfrag{mr}{$\bf{m_{\tilde{e}_R}} $}
\includegraphics[width=6.5cm, height=5.15cm]{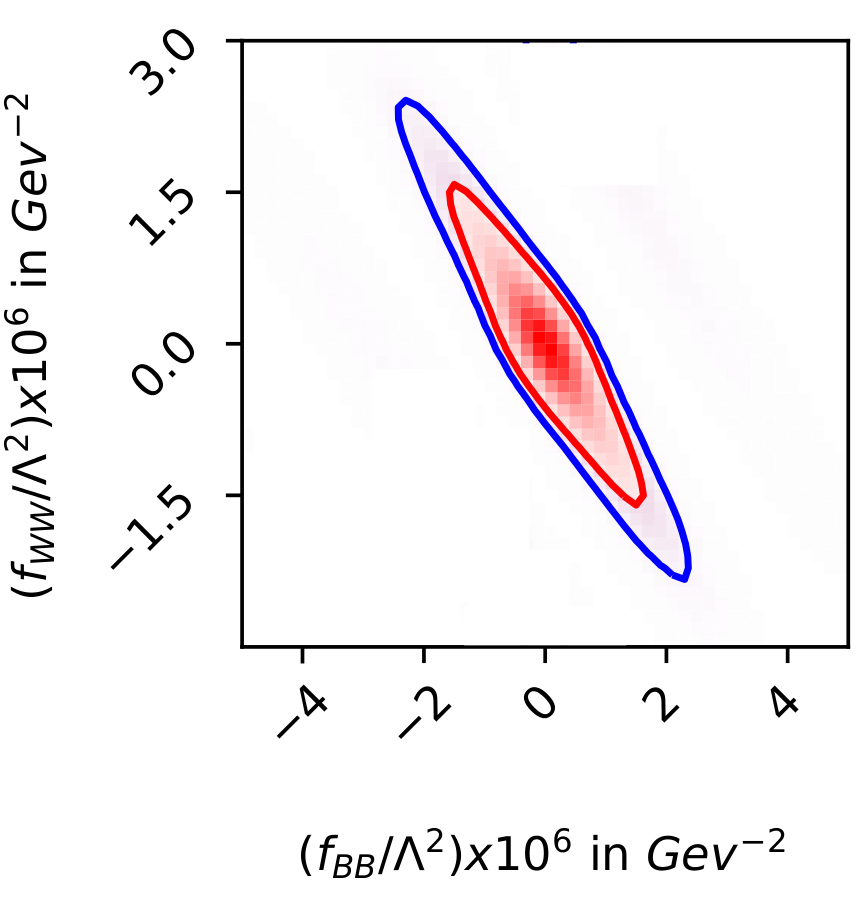}
%\caption{}
%\label{fig:mssm_br}
\end{minipage}
\caption{Allowed regions at $1\sigma$(red) and $2\sigma$(blue)
  levels in the parameter space of scale factors and dimension-6
  couplings in Lepton-specific 2HDM}
\label{figure11}
\end{figure}   

\begin{figure}[!htb]
\begin{minipage}[!htb]{0.45\linewidth}
\centering
%\vspace*{0.7cm}
%\psfrag{ml}{$\bf{m_{\tilde{e}_L}} $}
\includegraphics[width=6.5cm, height=5.15cm]{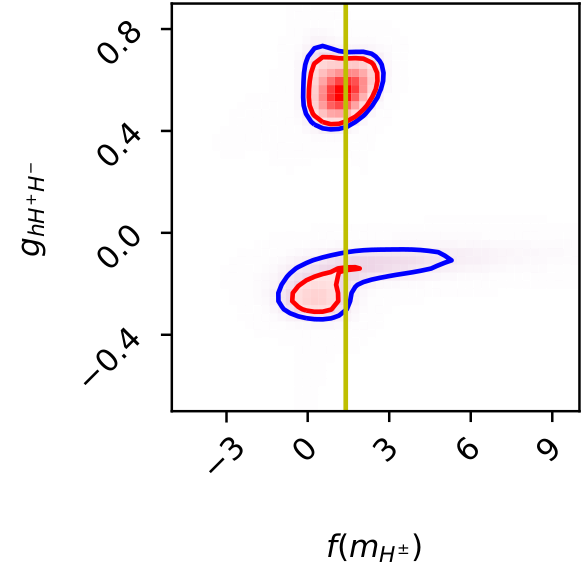}
%\caption{}
%\label{fig:mssm_ln}
\end{minipage}
\hspace{0.9cm}
\begin{minipage}[!htb]{0.45\linewidth}
\centering
%\psfrag{mr}{$\bf{m_{\tilde{e}_R}} $}
\includegraphics[width=6.5cm, height=5.15cm]{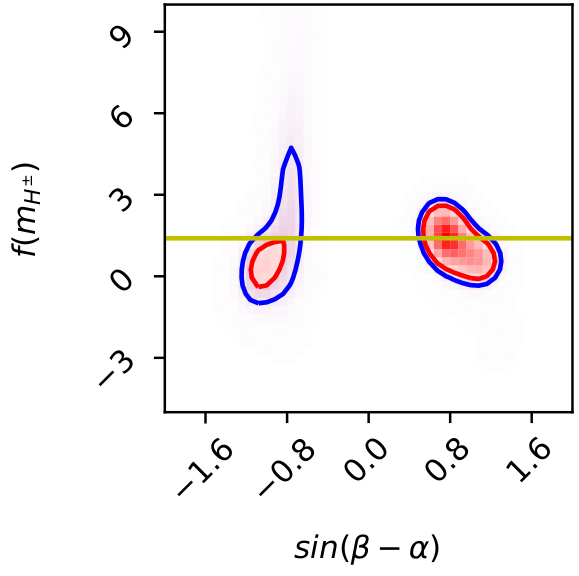}
%\caption{}
%\label{fig:mssm_br}
\end{minipage}
\caption{Allowed regions at $1\sigma$(red) and $2\sigma$(blue)
  levels in the parameter space of $f(m_{H^{\pm}})-g_{hH^+H^-}$ (left) and  $\sin(\beta-\alpha)-f(m_{H^{\pm}})$ (right) in Lepton-specific 2HDM}
\label{figure12}
\end{figure}

Our model independent 8-dimensional parameter space shrinks again to a
6-dimensional parameter space spanned by $\kappa_v, \kappa_t,
f_{BB}, f_{WW}, f_B, f_W$. The allowed parameter space
at $1\sigma$ and $2\sigma$ levels is shown in Figure~\ref{figure11}
(left) in $\kappa_v$-$\kappa_t$ plane and in Figure~\ref{figure11}
(right) in $f_{BB}$-$f_{WW}$ plane, from which one sets about
translating these limits to those on the model parameter space.

\begin{eqnarray}
\kappa_v &=& \sin(\beta - \alpha) \\ \kappa_t &=& \frac{\cos
  \alpha}{\sin \beta} = \sin(\beta - \alpha) - \cos(\beta - \alpha)
\tan \beta
\end{eqnarray}

Equating the loop contribution to  $h \rightarrow
\gamma\gamma$ from both the approaches, we plot the $1\sigma$ and
$2\sigma$ contours in the $f(m_{H^{\pm}})$-$g_{hH^+H^-}$ plane and
$\sin (\beta - \alpha$)-$f(m_{H^{\pm}})$ plane in Figure~\ref{figure12}
(left) and (right) respectively. Only the regions
corresponding to $f(m_{H^{\pm}}) > 0$ on the right-hand side of the
vertical line and above the horizontal line are allowed,
if one demands $f(m_{H^\pm}) \ge 80$ GeV.

\subsubsection{Flipped 2HDM}

Here the up-type quarks and leptons couple to the same
doublet while the other doublet couples to down-type quarks. The Yukawa
Lagrangian becomes

\begin{equation}
\label{flipped}
{\cal L}_{Yukawa} = y^1_{ij}\bar Q_{iL} \Phi_1 d_{jR} + y^2_{ij}\bar
Q_{iL} {\bar{\Phi}}_2 u_{jR} + y^5_{ij}\bar L_{iL} \Phi_2 e_{jR}
\end{equation}
which is achieved by imposing  $\Phi_1
\rightarrow -\Phi_1$ and $d_R \rightarrow -d_R$. 
It is straightforward to show that the scale factors
with respect to SM Higgs couplings to fermions and gauge bosons then
become:

\begin{eqnarray}
C_{h t \bar t} = \frac{\cos \alpha}{\sin \beta} \times C_{h t \bar
  t}^{\text SM} \nonumber \\ C_{h b \bar b} = -\frac{\sin \alpha}{\cos
  \beta} \times C_{h b \bar b}^{\text SM} \nonumber \\ C_{h \tau \bar
  \tau} = \frac{\cos \alpha}{\sin \beta}\times C_{h \tau \bar
  \tau}^{\text SM} \nonumber \\ C_{hVV} = \sin(\beta - \alpha) \times
C_{hVV}^{\text SM}
\label{mod4}
\end{eqnarray}
and the scaling factors are

\begin{eqnarray}
\kappa_t &=& \kappa_{\tau}  \\
\kappa_{b} &=& \frac{\kappa_t^2 - 2 \kappa_t \kappa_v +1}{\kappa_t - \kappa_v}
\end{eqnarray}

\begin{figure}[!htb]
\begin{minipage}[!htb]{0.45\linewidth}
\centering
%\vspace*{0.7cm}
%\psfrag{ml}{$\bf{m_{\tilde{e}_L}} $}
\includegraphics[width=6.5cm, height=5.15cm]{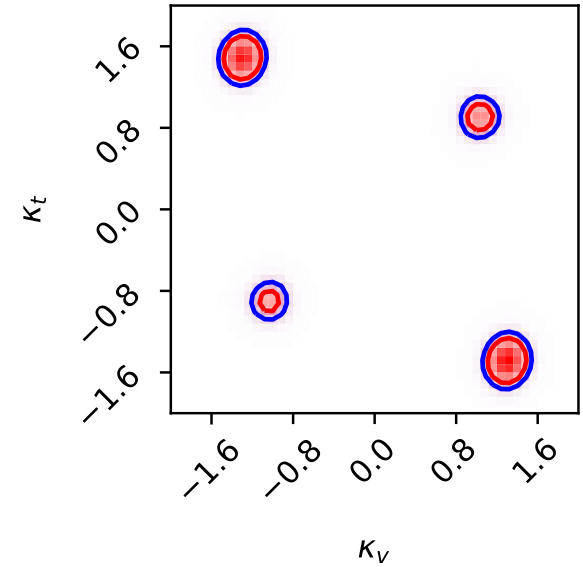}
%\caption{}
%\label{fig:mssm_ln}
\end{minipage}
\hspace{0.9cm}
\begin{minipage}[!htb]{0.45\linewidth}
\centering
%\psfrag{mr}{$\bf{m_{\tilde{e}_R}} $}
\includegraphics[width=6.5cm, height=5.15cm]{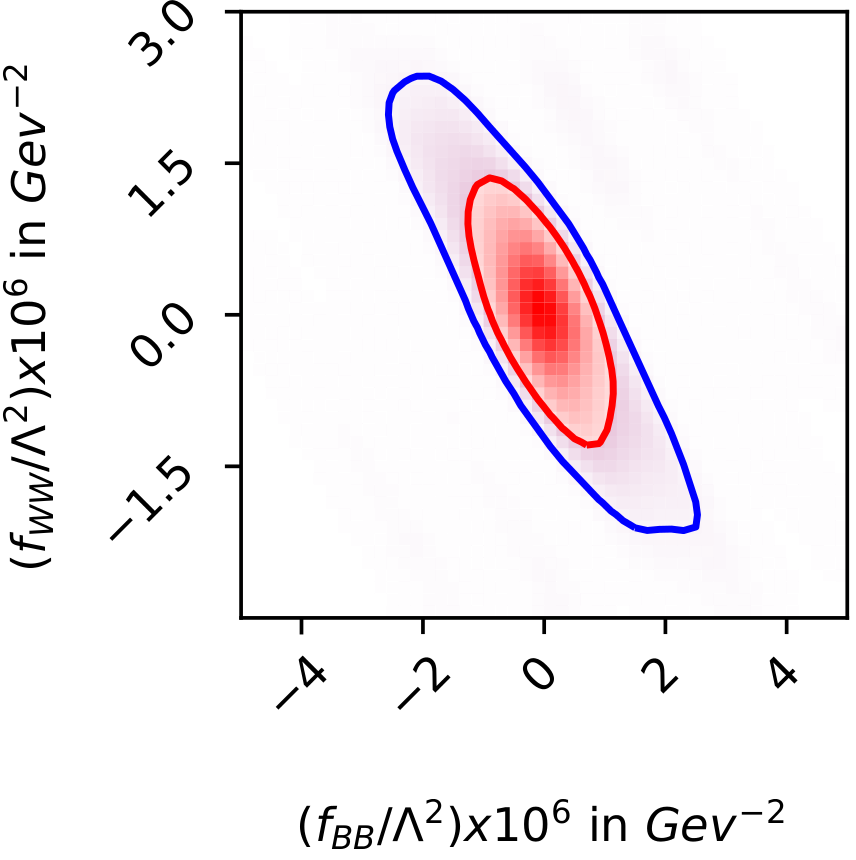}
%\caption{}
%\label{fig:mssm_br}
\end{minipage}
\caption{Allowed regions at $1\sigma$(red) and $2\sigma$(blue) levels
   in the parameter space of scale factors and dimension-6
  couplings in Flipped 2HDM}
\label{figure13}
\end{figure}

Here our 6-dimensional parameter space are comprised of $\kappa_v,
\kappa_t, f_{BB}, f_{WW}, f_B, f_W$ .  The $1\sigma$ and $2\sigma$
regions are shown in Figure~\ref{figure13} (left) in the
$\kappa_v$-$\kappa_t$ plane and in Figure~\ref{figure13} (right) in
$f_{BB}$-$f_{WW}$ plane, from which one obtains the corresponding
regions in the model parameter space, following a strategy similar to
the one used previously.

\begin{eqnarray}
\kappa_v &=& \sin(\beta - \alpha) \\ \kappa_t &=& \frac{\cos
  \alpha}{\sin \beta} = \sin(\beta - \alpha) - \cos(\beta - \alpha)
\tan \beta
\end{eqnarray}

\begin{figure}[!htb]
\begin{minipage}[!htb]{0.45\linewidth}
\centering
%\vspace*{0.7cm}
%\psfrag{ml}{$\bf{m_{\tilde{e}_L}} $}
\includegraphics[width=6.5cm, height=5.15cm]{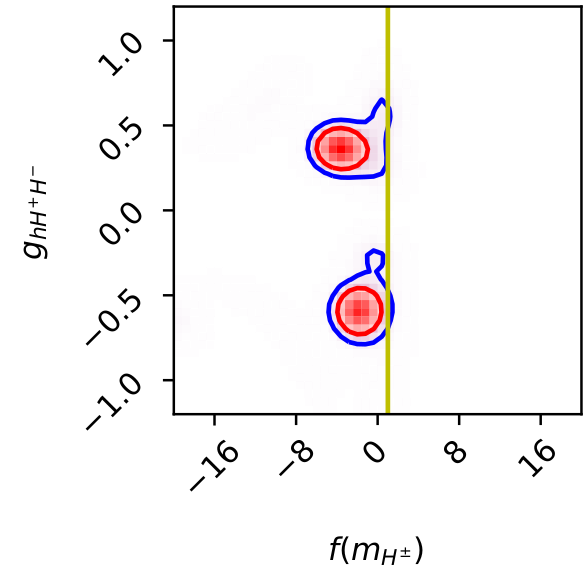}
%\caption{}
%\label{fig:mssm_ln}
\end{minipage}
\hspace{0.9cm}
\begin{minipage}[!htb]{0.45\linewidth}
\centering
%\psfrag{mr}{$\bf{m_{\tilde{e}_R}} $}
\includegraphics[width=6.5cm, height=5.15cm]{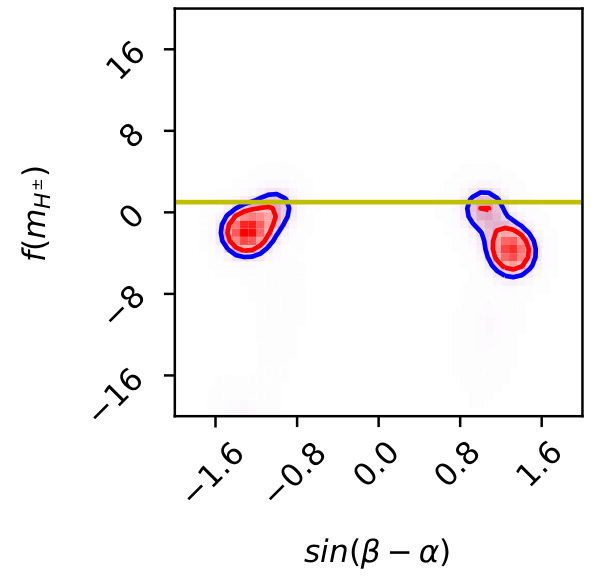}
%\caption{}
%\label{fig:mssm_br}
\end{minipage}
\caption{Allowed regions at $1\sigma$(red) and $2\sigma$(blue)
  levels in the parameter space of $f(m_{H^{\pm}})-g_{hH^+H^-}$ (left) and  $\sin(\beta-\alpha)-f(m_{H^{\pm}})$ (right) in Flipped 2HDM}
\label{figure14}
\end{figure}

One thus obtains $1\sigma$ and $2\sigma$ contours in the
$f(m_{H^{\pm}})$-$g_{hH^+H^-}$ plane and $\sin (\beta -
\alpha)$-$f(m_{H^{\pm}})$ plane in Figure~\ref{figure14} (left) and (right) respectively. Here, too, a lower limit on the charged Higgs mass
similar to that for Type-II 2HDM is applicable. 

%%----------------------------------------

\subsubsection{Comparison among various 2HDMs}

We next ask the question: while allowed regions
for the various 2HDM scenarios are identified as above from
available data, can some quantities be defined further, which
may act as differentiators among them? We are essentially occupied with 
one such quantity in the rest of this subsection.

\begin{figure}[!hptb]
%\begin{minipage}[h!]{0.45\linewidth}
%\centering
%\vspace*{0.7cm}
%\psfrag{ml}{$\bf{m_{\tilde{e}_L}} $}
\includegraphics[width=7.5cm, height=7cm]{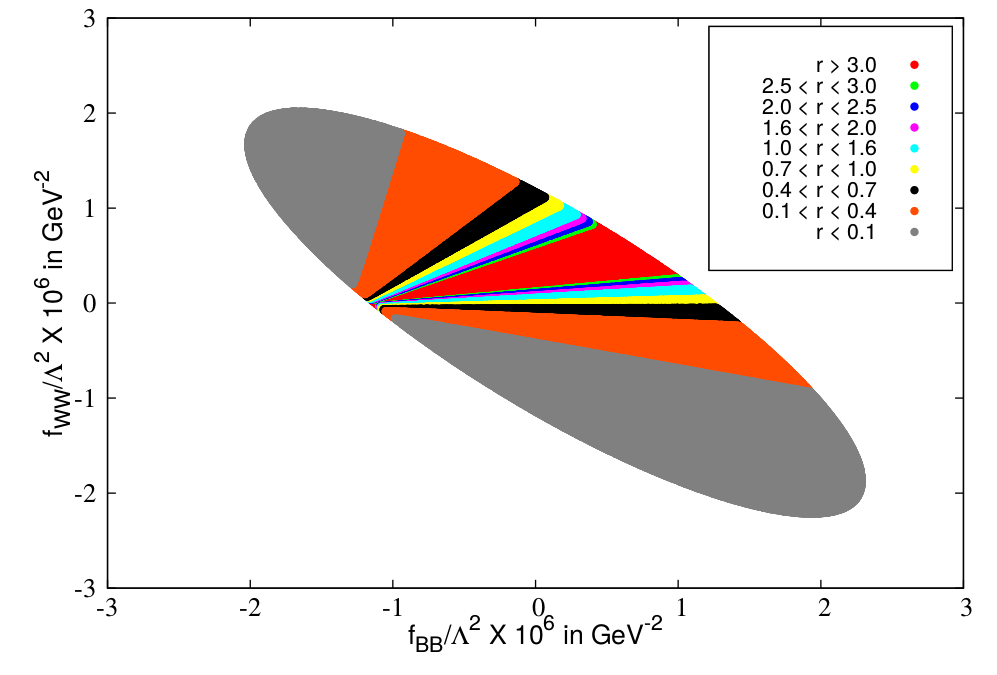}
%\caption{}
%\label{fig:mssm_ln}
%\end{minipage}
\hspace{0.9cm}
%\begin{minipage}[h!]{0.45\linewidth}
%\centering
%\vspace*{0.7cm}
%\psfrag{mr}{$\bf{m_{\tilde{e}_R}} $}
\includegraphics[width=7.5cm, height=7cm]{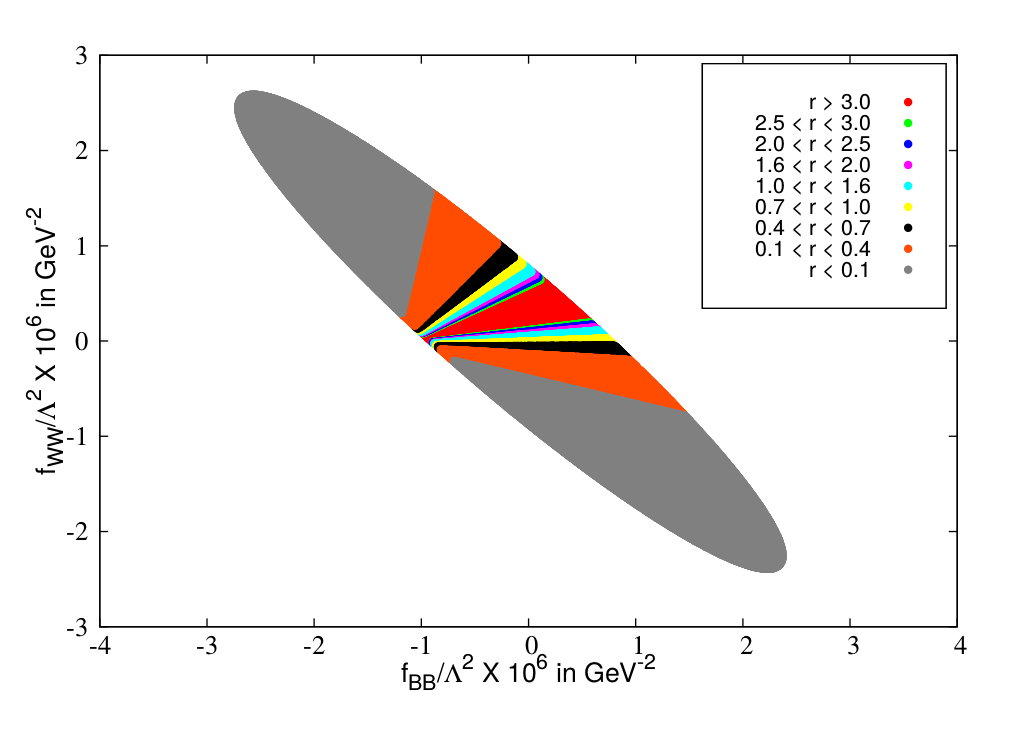} \\
%\caption{}
%\label{fig:mssm_br}
%\end{minipage}
%%%%%%%%%%%%%%%%%%%%%%%%%%%%%%%%%
%\begin{minipage}[h!]{0.45\linewidth}
%\centering
\vspace*{0.02cm}
%\psfrag{ml}{$\bf{m_{\tilde{e}_L}} $}
\includegraphics[width=7.5cm, height=7cm]{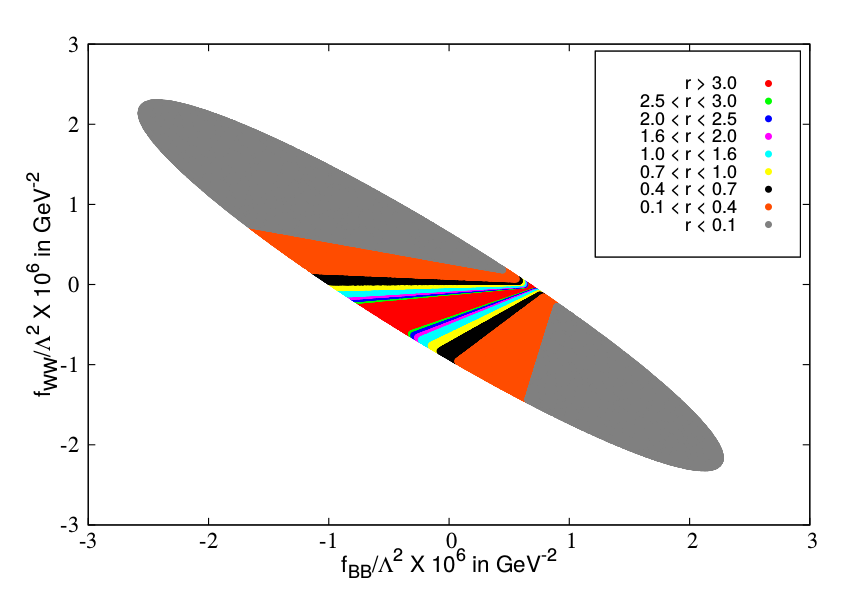}
%\caption{}
%\label{fig:mssm_ln}
%\end{minipage}
\hspace{0.9cm}
%\begin{minipage}[h!]{0.45\linewidth}
%\centering
%\vspace*{0.7cm}
%\psfrag{mr}{$\bf{m_{\tilde{e}_R}} $}
\includegraphics[width=7.5cm, height=7cm]{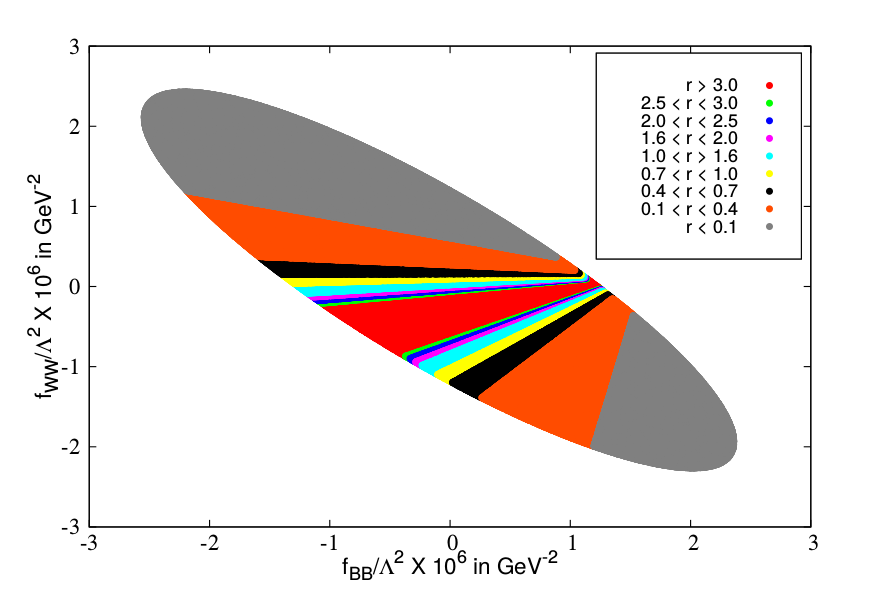}
%\caption{}
%\label{fig:mssm_br}
%\end{minipage}
%%%%%%%%%%%%%%%%%%%%%%%%%%%%%%%%%%%%%%%%
\caption{The ratio $r$ in $f_{BB}$-$f_{WW}$ plane for Type I(top left),
  Type II(top right), Lepton-specific(bottom left) and Flipped
  2HDM(bottom right). The ellipses denote the regions allowed at 2$\sigma$ level.}
\label{figure15}
\end{figure}   

\begin{figure}[h!]
\begin{minipage}[h!]{0.45\linewidth}
\centering
%\vspace*{0.7cm}
%\psfrag{ml}{$\bf{m_{\tilde{e}_L}} $}
\includegraphics[width=7.5cm, height=6cm]{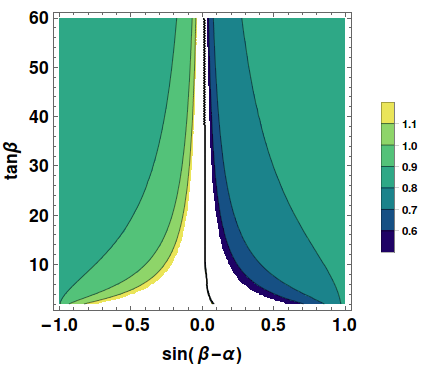}
%\caption{}
%\label{fig:mssm_ln}
\end{minipage}
\hspace{1.35cm}
\begin{minipage}[h!]{0.45\linewidth}
\centering
%\vspace*{0.7cm}
%\psfrag{mr}{$\bf{m_{\tilde{e}_R}} $}
\includegraphics[width=7.5cm, height=6cm]{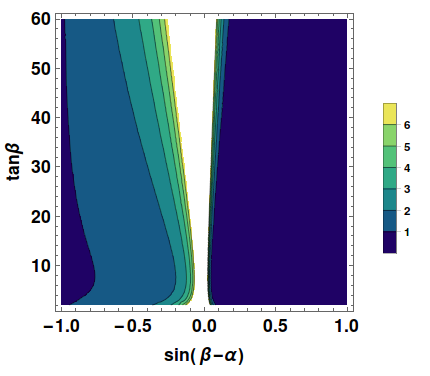}
%\caption{}
%\label{fig:mssm_br}
\end{minipage}
%%%%%%%%%%%%%%%%%%%%%%%%%%%%%%%%%
\begin{minipage}[h!]{0.45\linewidth}
\centering
\vspace*{0.7cm}
%\psfrag{ml}{$\bf{m_{\tilde{e}_L}} $}
\includegraphics[width=7.5cm, height=6cm]{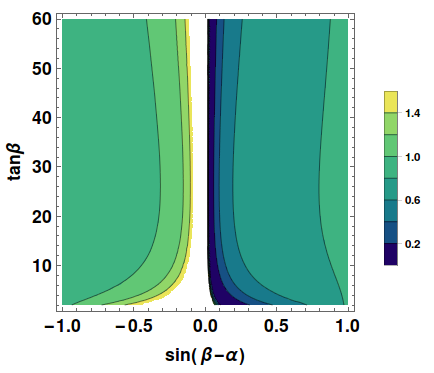}
%\caption{}
%\label{fig:mssm_ln}
\end{minipage}
\hspace{1.35cm}
\begin{minipage}[h!]{0.45\linewidth}
\centering
\vspace*{0.7cm}
%\psfrag{mr}{$\bf{m_{\tilde{e}_R}} $}
\includegraphics[width=7.5cm, height=6cm]{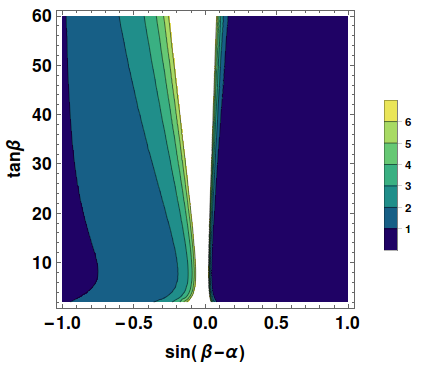}
%\caption{}
%\label{fig:mssm_br}
\end{minipage}
%%%%%%%%%%%%%%%%%%%%%%%%%%%%%%%%%%%%%%%%
\caption{The ratio $r$ in $\sin(\beta - \alpha$)-$\tan \beta$ plane for
  Type I(top left), Type II(top right), Lepton-specific(bottom left)
  and Flipped 2HDM(bottom right), where $m_{H^{\pm}}$=600 GeV.}
\label{figure16}
\end{figure}   

 With this in view, we consider the ratio 

\begin{equation} 
r= \frac{\mu_{\gamma\gamma}}{\mu_{Z\gamma}}
\end{equation}
% \frac{\sigma(pp \rightarrow h)\text{Br}(h \rightarrow
%     \gamma\gamma)}{\sigma(pp \rightarrow h )\text{Br}(h \rightarrow
%     Z\gamma)}/\frac{\sigma(pp \rightarrow h)\text{Br}(h \rightarrow
%     \gamma\gamma)}{\sigma(pp \rightarrow h)\text{Br}(h \rightarrow
%     Z\gamma)}_{SM} 
    
\noindent
which essentially reduces to
 $\frac{\Gamma(h \rightarrow \gamma \gamma)}{\Gamma(h \rightarrow Z
   \gamma)}   /\frac{{\Gamma(h \rightarrow \gamma \gamma)}_{SM}}{{\Gamma(h
   \rightarrow Z \gamma)}_{SM}}$, when all the major channels
 of Higgs production are summed over. We show color-coded regions
 in the $f_{BB}$-$f_{WW}$ plane for the four aforesaid types of 2HDM
 in Figure~\ref{figure15}. As can be seen there,  
 $r$ can range form values below 0.1 to those as large as 3, consistently
 with the current data. 
As progressively accumulating data shrink
the $2\sigma$ regions, thereby causing the ellipses to become smaller,
the observation of this ratio can narrow down the choice among
different models at least partially.
One rather striking observation here is the opposite directions
of convergence of the color-coded regions for Types I and II as against
those for the Lepton-specific and Flipped cases. It should also be noted that
in general the allowed regions for Type-I and Flipped category are slightly
bigger than those of Type-II and Lepton-specific case. While drawing these plots the scale factors $\kappa$'s have been kept at their best-fit values and the dimension-6 couplings $f_B$ and $f_W$ are assumed to be zero.

 Color-coded regions corresponding to different ranges of $r$ in the
 $\sin(\beta - \alpha)$-$\tan \beta$ plane are shown in
 Figure~\ref{figure16}. 
 In case of Type I 2HDM, for $\sin(\beta -
 \alpha) \approx +1$(the so-called `alignment limit'~\cite{Bernon:2015qea} indicates $\sin(\beta -
 \alpha) \approx \pm 1$), and low $\tan \beta$, the top loop contribution
 to $h \rightarrow \gamma \gamma$ increases and interferes
 destructively with the $W$-loop contribution. The charged Higgs loop also
 interferes destructively with the $W$ loop. These contributions thus 
 collectively reduce the decay width as compared to
 its SM value. 

 On the other hand, in the case of $h \rightarrow Z\gamma$,
 the top and charged Higgs loops interfere constructively with
 the $W$-loop. Therefore, despite the marginal fall in the $W$-loop 
 contribution from the SM value, driven by  $\sin^2(\beta-\alpha)$, 
 such constructive interference partially compensates it. Hence for
 positive $\sin(\beta-\alpha$)  the ratio $r$ is always less than
 unity. 

 In the region with $\sin(\beta-\alpha) \approx -1$, 
 $\Gamma(h \rightarrow \gamma \gamma)$ has a small suppression over the
 entire chosen range of $\tan \beta$. On the other hand, the top
 loop contribution to $h \rightarrow Z\gamma$ there decreases more
 significantly at small $\tan \beta$. Coupled with the effect of
 constructive interference of the top, $W$ and charged Higgs loops, this 
 reduces the overall $h \rightarrow Z\gamma$ decay width from its SM
 value. Hence the ratio $r$ in this case can be greater than unity for
 low $\tan \beta$, as can be seen in Figure~\ref{figure16} (top left). 
 The Lepton-specific scenario shows a trend similar to Type I, as is clear
 from Figure~\ref{figure16} (bottom left), since the quark couplings to
 Higgs are same in both the cases.

 Figure~\ref{figure16} shows that, for
 Type II and Flipped 2HDM, the ratio is always $<1$ if
 $\sin(\beta-\alpha)$ is not much different from +1. But in the wrong-sign
 region where $\sin(\beta-\alpha) \approx -1$ which gives rise to
 a positive sign in the coupling of Higgs with down-type quarks, $r$
 can be greater than 1 for large enough $\tan \beta$. For such $\tan
 \beta$, the bottom loop contribution to $h \rightarrow \gamma \gamma$
 increases. In case of positive $\sin (\beta - \alpha$) this loop
 interferes destructively with the $W$ loop and constructively 
 for negative $\sin (\beta - \alpha)$. The top and charged
 Higgs loops interfere destructively with the $W$ loop in both these
 cases. Hence the decay width of $h \rightarrow \gamma \gamma$ can be
 greater than its SM value for negative sign of $\sin (\beta - \alpha$). 
 For $h \rightarrow Z \gamma$  in the positive $\sin(\beta -
 \alpha$) region, the bottom loop constructively interferes with the
 $W$ and top loop, where in the negative $\sin(\beta - \alpha$) regime, the
 bottom loop destructively interferes with the $W$ and top loops. The
 charged Higgs loop contribution is much smaller compared to the rest
 of the contributions. At large $\tan \beta$ the contribution
 increases and therefore the $h \rightarrow Z\gamma$ decay width
 decreases below its SM value for negative $\sin(\beta - \alpha$) and
 increase above that SM value for positive $\sin(\beta - \alpha$). 
 Thus in general the ratio $r$ can be greater than unity for large
 $\tan \beta$ and for the wrong sign region for Type II 2HDM, as can
 be seen from Figure~\ref{figure16}(top right). The Flipped 2HDM scenario,
 too follows the same trend, as is evident from
 Figure~\ref{figure16}(bottom right) . 
 
 We have checked the validity of the aforementioned analysis over a wide range of charged Higgs masses. In Figure~\ref{figure16} we present the results for a charged Higgs mass of 600 GeV. But the results remain unaltered even for $m_{H^{\pm}} \sim {\cal O}$(TeV).

\subsection{Higgs triplet models}

%~\cite{1601.05217}
As already mentioned, a pertinent issue for SM has been 
neutrino masses and mixing. A popular 
model that explains these, by introducing an extension of
the EWSB sector, is the Type-II see-saw
mechanism~\cite{delAguila:2008ks,Arhrib:2009mz,Akeroyd:2005gt,Akeroyd:2009hb,Akeroyd:2010je,Akeroyd:2011zza,Chakrabarti:1998qy,Cheung:2002gd,Chiang:2012dk,Kanemura:2013vxa,
kang:2014jia,Kanemura:2014goa,Kanemura:2014ipa,Chen:2014qda,Han:2015hba,Han:2015sca,Mitra:2016wpr,Ghosh:2017pxl}. This scenario allows one 
to dispense with right-handed
neutrinos, having instead 
one or more scalar triplets with hypercharge 2. Rather strong constraints
arising from the $\rho$-parameter
apply on the vev of the triplet(s), restricting them to
values ${\cal{O}}\sim$ 1 GeV. Moreover, a bid to unification via left-right symmetry
~\cite{Pati:1974yy,Mohapatra:1974hk,Mohapatra:1974gc} makes use of
such triplet scalars. 

The bi-large neutrino mixing pattern, the confirmation of the 
non-vanishing mixing element $\theta_{13}$, and the various suggested 
mass hierarchies prompt one to set up neutrino mass
models which try to connect mass eigenvalues with the mixing angles,
thereby adding predictiveness to the theoretical formulation. 
Such predictiveness is enhanced if the neutrino mass matrix in the
flavor basis contains some zero entries, ensured by the imposition
of additional symmetries. However, a single triplet
fails to make Type II seesaw scenarios consistent with some
widely used zero texture models, in particular, those with two-zero
texture. On the contrary, such texture can be reproduced if one has
two triplets instead of one, thereby suggesting a scenario whose
phenomenological implications have been explored in recent 
works~\cite{Chaudhuri:2016rwo} and the collider signatures of this model have been studied in~\cite{Ghosh:2018jpa}. With this in mind, we analyze below models
both with one and two scalar triplet in the light of the recent Higgs data
and the contribution of charged scalar loops (including those from
doubly charged scalars) embodied in dimension-6 operators.

\subsubsection{A single-triplet model}

Here a bi-doublet $\Delta$ is added in the Higgs sector. Therefore in
addition to the SM Higgs state we have three more Higgs states, namely,
doubly charged $\Delta^{++}$, singly charged $\Delta^{+}$ and neutral $\Delta^0$
ones.

\begin{equation}\label{triplet} 
\Delta = 
\left(\begin{array}{cc}
\Delta^+ & \sqrt{2}\Delta^{++} \\ \sqrt{2}\Delta^0 & -\Delta^+ 
\end{array}\right).
\end{equation}

The vev of the doublet $\phi$ and triplet $\Delta$ can be expressed as
\begin{equation}\label{vev}
{\braket{\phi}}_0 =
\frac{1}{\sqrt{2}} \left(\begin{array}{c} 0 \\ v \end{array}\right)
\quad \mbox{and} \quad
{\braket{\Delta}}_0  =
\left(\begin{array}{cc} 0 & 0 \\ w & 0 \end{array}\right),
\end{equation}

The scalar potential in its most general form can be written as 

\begin{eqnarray}
V(\phi,\Delta)\  = \hphantom{+} a\, \phi^\dagger\phi + \ \frac{b}{2}
\text{Tr}(\Delta^\dagger\Delta)+ (\phi^\dagger\phi)^2   
+ \ \frac{d}{4}\left( \text{Tr}(\Delta^\dagger\Delta) \right)^2 \nonumber \\ +
\ \frac{e - h}{2} \phi^\dagger\phi \text{Tr}(\Delta^\dagger\Delta)  
+  \ \frac{f}{4} \text{Tr}(\Delta^\dagger\Delta^\dagger)\text{Tr}(\Delta\Delta)
\nonumber \\ 
+ \ h \phi^\dagger \Delta^\dagger \Delta \phi + \left(t \phi^\dagger
\Delta \tilde{\phi} + 
\mbox{H.c.} \right),
\end{eqnarray}
where $\tilde{\phi} \equiv i\tau_2 \phi^\ast$.

We neglect CP-violation for simplicity. Thus
all the coefficients in the scalar potential are assumed to be
real. To ensure that  EWSB  is driven predominantly by
the doublet state, certain restrictions are imposed on the parameters. 
The choice $a<0$, $b>0$, takes care of this. In
addition, $w \ll v$, as required by the $\rho$-parameter
constraint.  The doublet-triplet mixing should be small in general. 
In addition, we have confined ourselves to cases where 
all quartic couplings are perturbative. These requirements guide
us to choose the parameter space in the following manner:
\begin{equation}\label{oomagn}
a,\: b \sim v^2; \quad
c,\: d,\: e,\: f,\: h \sim 1; \quad 
|t|\ll v.  
\end{equation}

\begin{figure}[!hptb]
\centering
%\vspace*{0.7cm}
%\psfrag{ml}{$\bf{m_{\tilde{e}_L}} $}
\includegraphics[width=6.8cm, height=5.3cm]{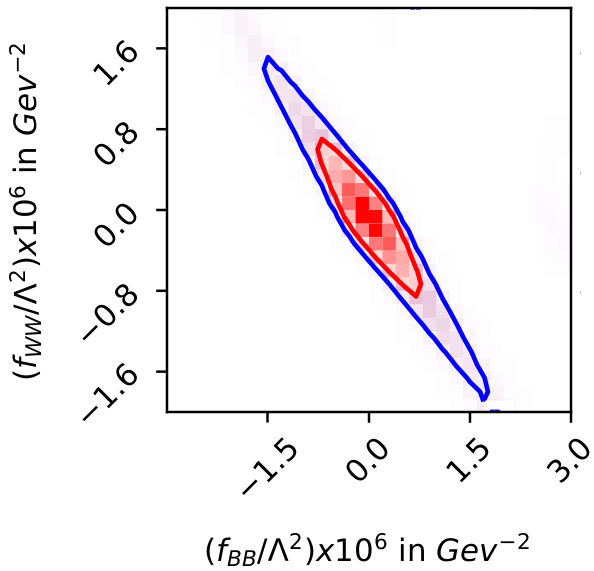}
\hspace{0.4cm}
\includegraphics[width=6.8cm, height=5.3cm]{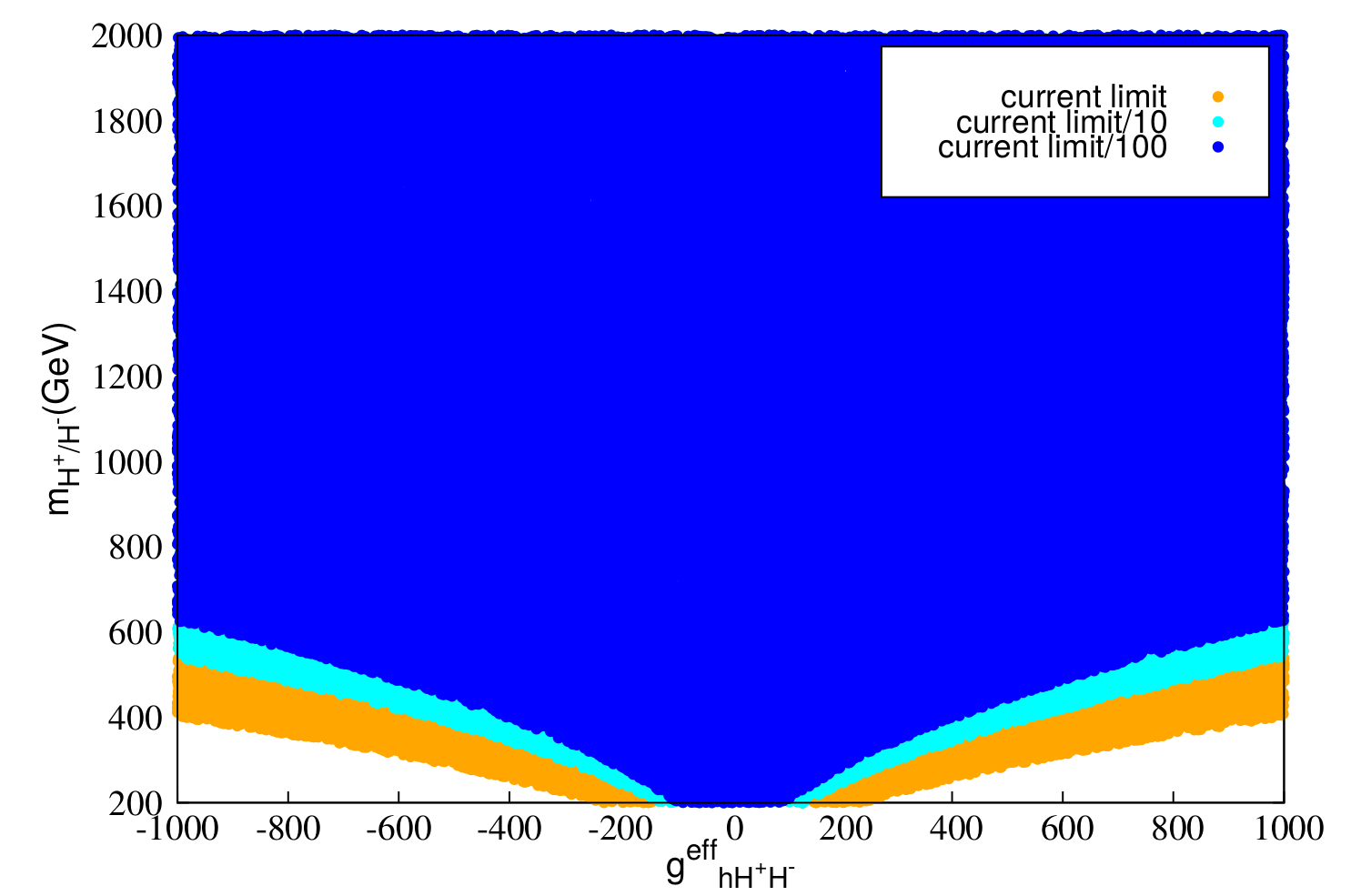} \\
\includegraphics[width=6.8cm, height=5.3cm]{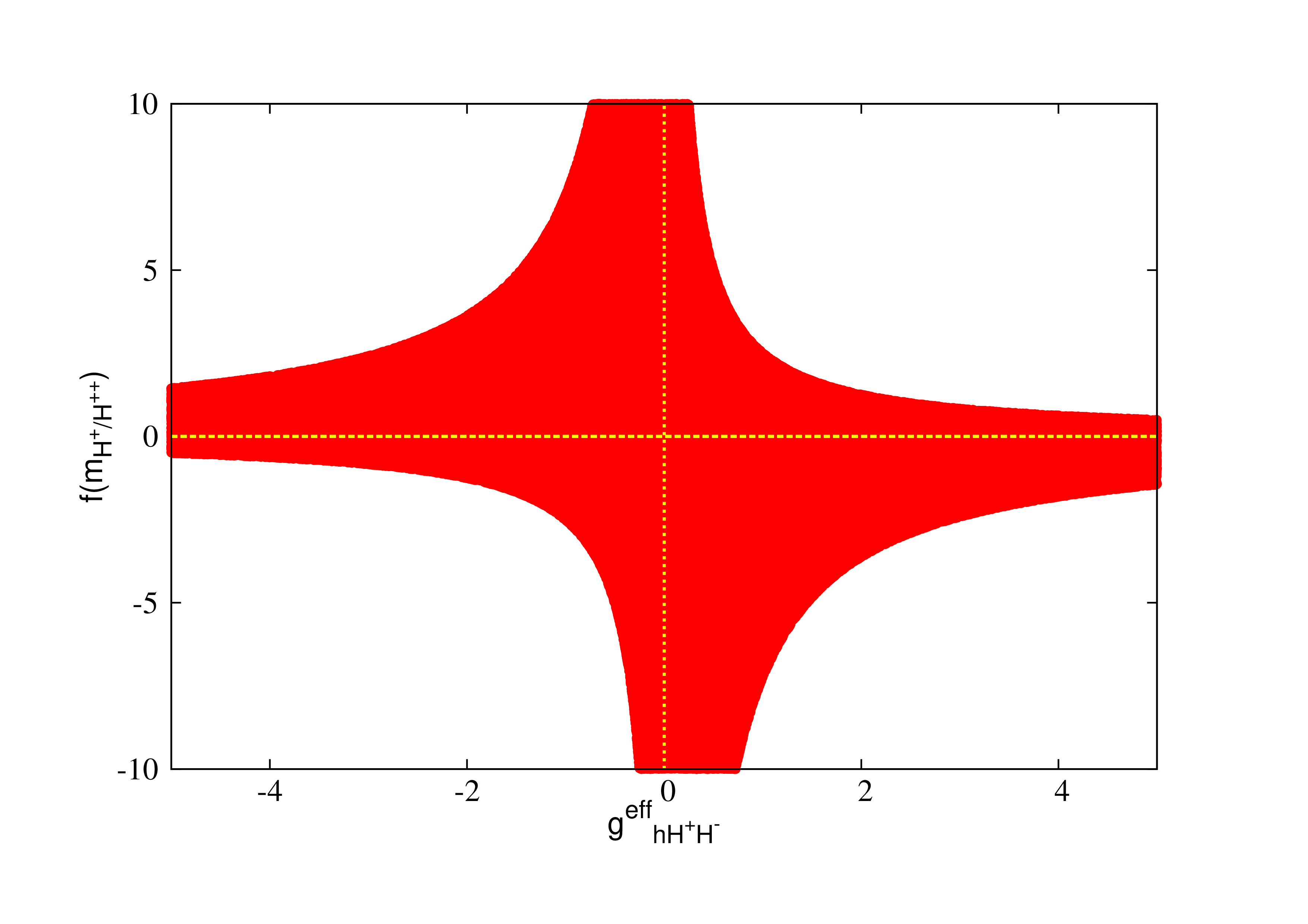}
%\caption{}
%\label{fig:mssm_ln}
%\end{minipage}
\caption{Allowed regions at $1\sigma$(red) and $2\sigma$(blue)
 levels in the parameter space of scale factors and dimension-6
  couplings in the triplet scenario}
\label{figure17}
\end{figure}   

After spontaneous symmetry breaking, we are left with five physical
states, namely, two neutral scalars (one SM-like among them and one
heavier), one each of  singly-and doubly-charged scalars, 
one pseudoscalar. SU(2) invariance of the Lagrangian plus the smallness
demands of  doublet-triplet mixing force masses of the triplet-dominated 
states to be nearly degenerate. In such a situation 
SM-like Higgs couplings to fermions and gauge bosons are practically
unaltered. Therefore we can assume 
\beq
\kappa_v \approx \kappa_t \approx \kappa_b \approx \kappa_{\tau}
\approx 1 
\eeq 
Both the charged Higgs states will contribute to the
loop-induced decays of the SM-like Higgs $h$, the contribution
of the doubly charged Higgs loop to the width being four times that of 
the singly charged one. The decay rates of $h\rightarrow \gamma\gamma$ and $h\rightarrow Z\gamma$ in single-triplet case are
given by~\cite{Chen:2013vi,Chen:2013dh,Akeroyd:2012ms}. We also discuss them in detail in Appendix~\ref{app2}.

Since the couplings of $h$ to fermions and gauge bosons are
practically unchanged, 
the one-loop contribution to $h \rightarrow
\gamma \gamma, Z\gamma$ via the top-and W-loops are 
same as in SM. The loops induced by
the singly-and doubly-charged scalars can, as before,
be directly related to the dimension-6 couplings. 

The 1$\sigma$ and 2$\sigma$ regions in the  $f_{BB}$-$f_{WW}$
plane can now be obtained on the assumption that all $\kappa$'s are 
unity. Based on this, the
$1\sigma$ and $2\sigma$ contours in the $f_{BB}$-$f_{WW}$ plane 
are found in Figure~\ref{figure17} (top left). 
These are then directly translated into limits on 
the mass of the degenerate heavy scalar bosons. 

One can define an effective coupling as $g_{hH^+H^-}^{eff} =
\tilde{g}_{hH^+H^-}+4 \tilde{g}_{hH^{++}H^{--}}$, which can be
directly extracted from Equation~\ref{hgg} to be found in the
Appendix~\ref{app2}. Figure~\ref{figure17} (top right) is the translation of the
2$\sigma$ region of Figure~\ref{figure17} (top left) into the parameter
space spanned by the triplet mass scale $m_{H^{\pm}/H^{\pm\pm}}$ and
the effective coupling $g_{hH^+H^-}^{eff}$. The yellow region (over and above the cyan and blue region) here is
allowed by the current limit on $f_{BB}$ and $f_{WW}$. The region in
cyan (over and above the blue region) will be allowed if the limit on $f_{BB}$ and $f_{WW}$ improves by
a factor of 10.  The blue region represents the corresponding
projection assuming 100-fold improvement of the same
limits. Figure~\ref{figure17} (bottom centre) maps the same limit to the
$f(m_{H^{\pm}/H^{\pm\pm}})$-$g_{hH^+H^-}^{eff}$ parameter space,
showing directly the allowed values of the resultant loop function in
this scenario.

\subsubsection{A model with two triplets}

Based on the motivations discussed above, we next take up a
scenario with two triplet scalars with $Y=2$, namely, $\Delta_1$, $\Delta_2$,
each of which can be expressed as a bi-doublet:
\begin{equation}\label{triplet2}
\Delta_1 = 
\left(\begin{array}{cc}
\delta_1^+ & \sqrt{2}\delta_1^{++} \\ \sqrt{2}\delta_1^0 & -\delta_1^+ 
\end{array}\right)
\quad \mbox{and} \quad
\Delta_2 = 
\left(\begin{array}{cc}
\delta_2^+ & \sqrt{2}\delta_2^{++} \\ \sqrt{2}\delta_2^0 & -\delta_2^+ 
\end{array}\right).
\end{equation}
The vevs of the scalar triplets are given by
\begin{equation}\label{vev2'}
\braket{\Delta_1}_0 =
\left(\begin{array}{cc} 0 & 0 \\ w_1 & 0 \end{array}\right) 
\quad \mbox{and} \quad
\braket{\Delta_2}_0 =
\left(\begin{array}{cc} 0 & 0 \\ w_2 & 0 \end{array}\right).
\end{equation}
The vev of the Higgs doublet is given by Eq.~(\ref{vev}).

\begin{figure}[!hptb]
\centering
\includegraphics[width=10.8cm, height=9.3cm]{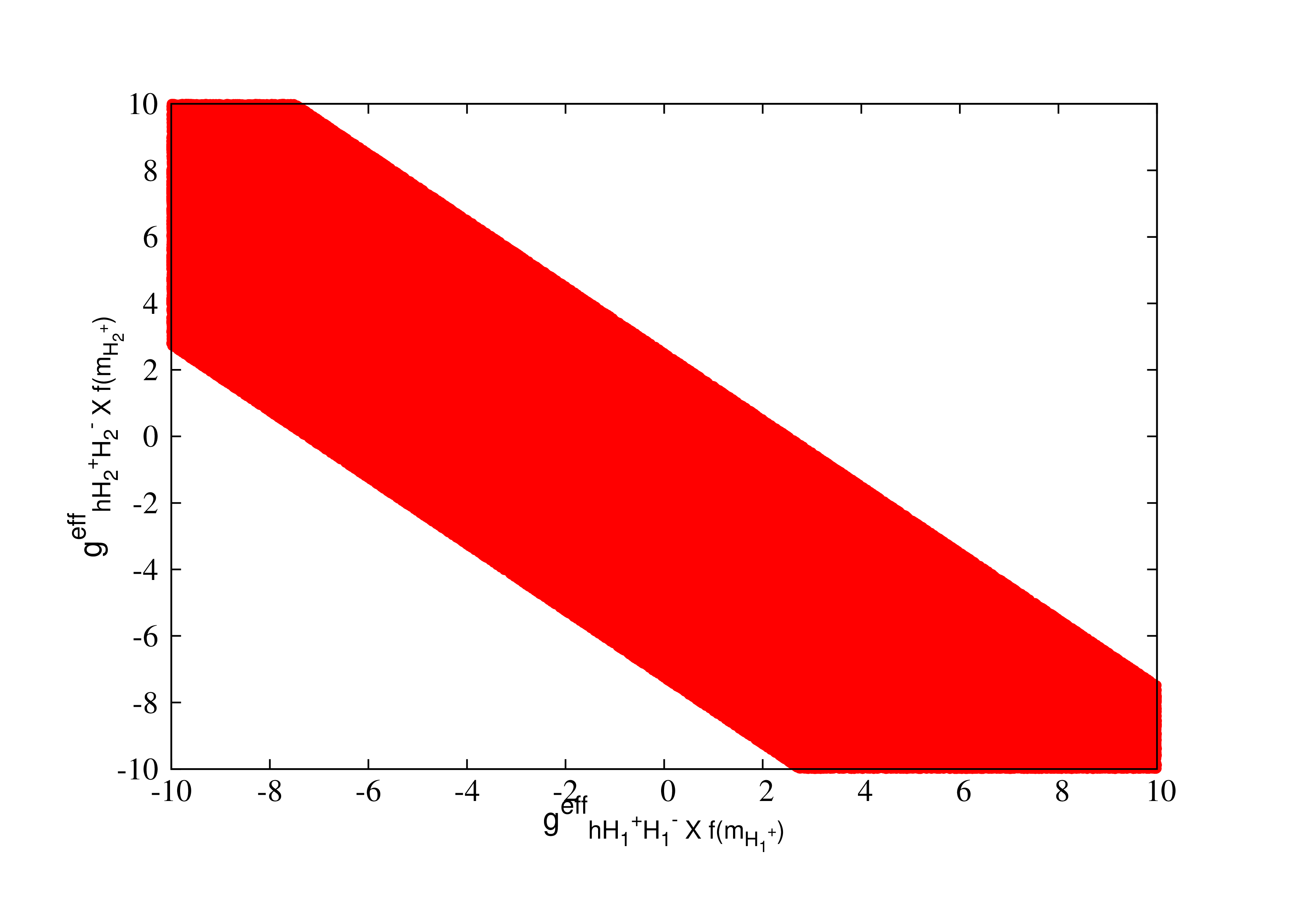}
\caption{The allowed region in the plane spanned by $g_{hH_1^+H_1^-}^{eff} \times f(m_{H_1^{\pm}})$ and $g_{hH_2^+H_2^-}^{eff} \times f(m_{H_2^{\pm}})$ at 2$\sigma$ level.}
\label{figure18}
\end{figure}

The scalar potential in this model involving $\phi$, $\Delta_1$
and $\Delta_2$ can be written as 
\begin{eqnarray}
V(\phi,\Delta_1,\Delta_2) =
\nonumber \\
&&
a \phi^\dagger\phi + 
\frac{1}{2}\, b_{kl} \text{Tr}(\Delta_k^\dagger \Delta_l)+ 
c (\phi^\dagger\phi)^2 + 
\frac{1}{4}\, d_{kl} \left( \text{Tr}(\Delta_k^\dagger\Delta_l) \right)^2
\nonumber \\
&&
+ \frac{1}{2}\,(e_{kl} - h_{kl})\,
\phi^\dagger \phi  \text{Tr} (\Delta_k^\dagger\Delta_l) +
\frac{1}{4}\,f_{kl}  
\text{Tr}(\Delta_k^\dagger\Delta_l^\dagger) \text{Tr}(\Delta_k\Delta_l)
\nonumber \\
&&
+ h_{kl}\, \phi^\dagger \Delta_k^\dagger \Delta_l \phi +
g \text{Tr}(\Delta_1^\dagger\Delta_2) \text{Tr}(\Delta_2^\dagger\Delta_1) + 
g' \text{Tr}(\Delta_1^\dagger\Delta_1) \text{Tr}(\Delta_2^\dagger\Delta_2) 
\nonumber \\
&&
+ \left( t_k\, \phi^\dagger \Delta_k \tilde{\phi} +
\mbox{H.c.} \right),
\label{pot}
\end{eqnarray}

The assumption that Higgs sector is CP-conserving prompts us to take
all the parameters in Equation~\ref{pot} as real. The parameters lie in
ranges very similar to those in the single-triplet case. Here, too,
the vev of the two triplets are kept on the order of a GeV. This
ensures negligible mixing between the doublet and the triplets,
again implying

\beq
\kappa_v \approx \kappa_t \approx \kappa_b \approx \kappa_{\tau} \approx 1
\eeq

%{\large \bf This part goes to the appendix:}
The decay rates for $h\rightarrow \gamma\gamma$ and $h\rightarrow Z \gamma$ in the two-triplet case are
given by~\cite{Chaudhuri:2016rwo}. We have also discussed them in Appendix~\ref{app2}.

%\subsubsection{Comparison between various triplets}

\begin{figure}[!hptb]
\centering
\includegraphics[width=10.8cm, height=9.3cm]{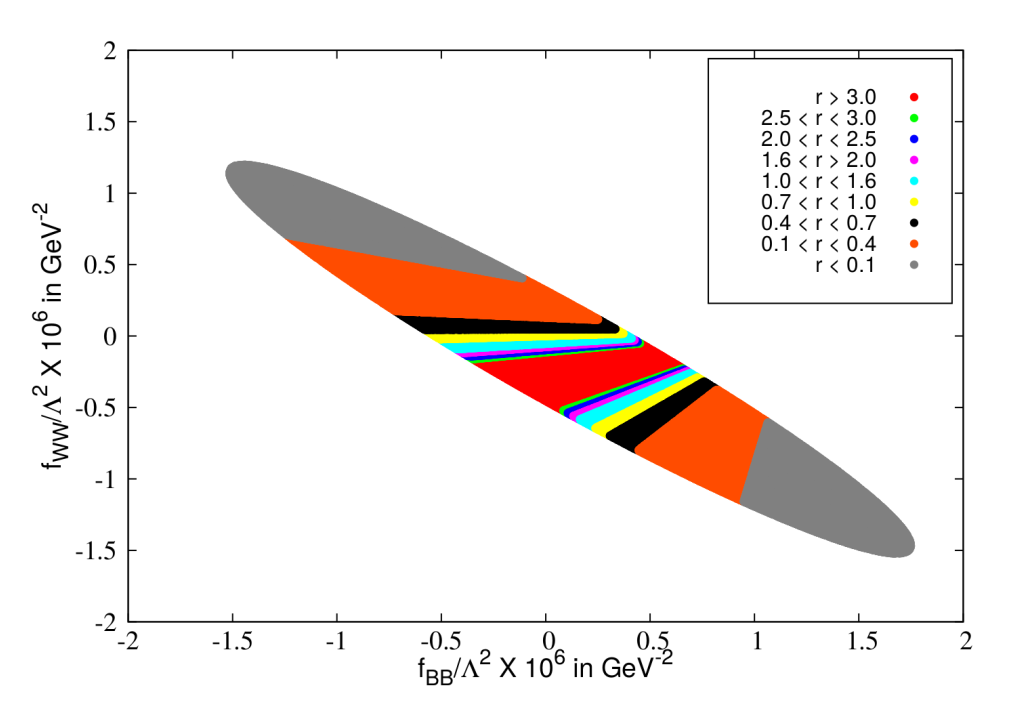}
\caption{The ratio $r$ in $f_{BB}$-$f_{WW}$ plane for triplet case. The ellipse denotes the region allowed at 2$\sigma$ level.}
\label{figure19}
\end{figure}

The allowed regions in the parameter space are illustrated in Figures~\ref{figure18} and ~\ref{figure19}, in the limit where the mixing between the two triplets is
negligible\footnote{This assumption in general has no justification;
  and is made here to enable us to express the singly- and
  doubly-charged scalar loop functions to have the same value in each
  triplet. If there is substantial mixing between them, one will have
  fours mass eigenstates altogether, and the loop functions will
  assume their values accordingly.}. The following relations then
hold: \beq m_{H_1^{\pm}} \approx m_{H_1^{\pm\pm}},\,\,\ m_{H_2^{\pm}}
\approx m_{H_2^{\pm\pm}} \eeq

and again we define effective couplings as in the singlet case:
\begin{eqnarray}
g_{hH_1^+H_1^-}^{eff} = \tilde{g_1}_{hH_1^+H_1^-}+4 \tilde{g_1}_{hH_1^{++}H_1^{--}} \\ 
g_{hH_2^+H_2^-}^{eff} = \tilde{g_2}_{hH_2^+H_2^-}+4 \tilde{g_2}_{hH_2^{++}H_2^{--}} 
\end{eqnarray}

In Figure~\ref{figure18}, we plot the allowed region in the
$g_{hH_1^+H_1^-}^{eff} \times f(m_{H_1^{\pm}})$ - $g_{hH_2^+H_2^-}^{eff} \times f(m_{H_2^{\pm}})$ plane at the 2$\sigma$ level,
following the same procedure as in the previous cases. In Figure~\ref{figure19} we
present the color-coded diagram of the ratio $r$ (as defined earlier)
in the $f_{BB}$-$f_{WW}$ plane for the triplet case. In this case the dimension-6 couplings $f_B$ and $f_W$ are assumed to be zero. If one compares this plot with
Figure~\ref{figure15}, one can see that the region corresponding to $r < 0.1$ is smaller
in case of triplet models than for any kind of  2HDM.

%\subsection{Comparison between models with doublets and triplets}

%The distinction between the various doublets and triplet models is in general possible in terms of the ratio $r$. 

\section{Conclusion}\label{summary}

We have explored the scope of new physics in the Higgs sector in the light of the
modified couplings with multiplicative scale factors
$\kappa_i$ as well as loop contributions parametrized by effective
dimension-6 operators(${\cal O}_{BB},{\cal O}_{WW},{\cal O}_{B}$ and ${\cal O}_{W}$). 
For the loop-induced decays $h \rightarrow \gamma \gamma, Z \gamma$, the SM-like loops are merely modified by the appropriate $\kappa$-parameters. In addition charged scalars may contribute in the loop induced processes such as $h \rightarrow \gamma \gamma , Z \gamma$. Thus these loop amplitudes contain effect of both the scale factors and the coefficients of the dimension-6 operators. We have imposed all existing experimental constraints and found the regions of parameter space, which are allowed at the 1 and 2$\sigma$ levels. While doing so we have taken into account the non-trivial correlation that may exist between various production mechanisms. Although there is no observation of $h \rightarrow Z \gamma$, the upper limit on this channel from current data puts considerable constraints on the parameter space spanned by $f_{BB}, f_{WW}, f_{B}$ and $f_W$.

We have tried next to establish connections between this model-independent effective field theory approach and specific new physics models with extended Higgs sector, namely various types of 2HDM and Higgs triplet models. We have observed that even when the heavy Higgs states in question are not much higher than the EWSB scale, it is possible to describe the new physics in terms of gauge-invariant effective operators. Thanks to this observation, we have made a correspondence between the model-independent and model-dependent approach and translated the limit obtained on the effective operators into those on specific model parameters. Thereafter, we have used the ratio between signal strengths in the $h \rightarrow \gamma \gamma$ and $h \rightarrow Z \gamma$ channel to show that this ratio helps in disentangling the effect of various types of new physics.

\newpage

\section*{Appendices}
\appendix
\section{Decay amplitudes in 2HDM}
\label{app1}

\noindent
The one-loop decay widths for $h \rightarrow \gamma
\gamma$ and $h \rightarrow Z \gamma$ under the framework of 2HDM can
be written as:
\begin{equation}
\label{hgamgam}
\Gamma(h \rightarrow \gamma \gamma) = \frac{\alpha^2 g^2}{1024 \pi^3}\frac{m_h^3}{m_W^2}|\sum_{i} N_{ci} R_i^{h} e_i^2 F_i|^2
\end{equation}
Where $i$ denotes spin index corresponding to spin-0(charged Higgs),
spin-1/2(top and bottom), and spin-1($W$), $e_i$ is the electric
charge in units of $e$.
\begin{eqnarray}
F_W &=& 2 + 3\tau + 3\tau(2-\tau)f^0(\tau) \nonumber \\
F_{t/b} &=& -2\tau[1+(1-\tau)f^0(\tau)]  \nonumber \\
F_{H^{\pm}} &=& \tau[1-\tau f^0(\tau)] 
\label{FWtop}
\end{eqnarray}
and 
\beq
\tau = \frac{4 m_i^2}{m_h^2}
\label{tau}
\eeq where $m_i$ is the mass of the particle running in the loop.

\beq f^0(\tau) = \left(\begin{array}{c} [\sin^{-1}
    (\sqrt{1/\tau)}]^2,~~~~~~~~~~\text{if}~~\tau \ge 1,
  \\ -\frac{1}{4}[\text{ln}(\eta_{+}/\eta_{-}) - i
    \pi]^2,~~\text{if}~~\tau <1
 \end{array}\right)
\label{ftau}
\eeq
%-\frac{1}{4}[ln(\eta_+/\eta_-) - i \pi]^2, if \tau <1 \right)

\begin{equation}
\label{hZgamma}
\Gamma(h \rightarrow Z \gamma) = \frac{\alpha^2 g^2}{512
  \pi^3}\frac{m_h^3}{m_W^2}\left(1-\frac{m_Z^2}{m_h^2}\right)^3|\sum_{i}
N_{ci} R_i^{h} A_i|^2
\end{equation}
The index i carries the same meaning as mentioned previously.

\begin{eqnarray}
A_W &=& -\cot \theta_W \left(4(3 - \tan^2 \theta_W)
I_2(\tau_W,\lambda_W)+ \left[\left(1+ \frac{2}{\tau_W}\right) \tan^2
  \theta_W - \left(5 +
  \frac{2}{\tau_W}\right)\right]I_1(\tau_W,\lambda_W) \right)
\nonumber \\ A_{f} &=& \frac{ -2 e_{f} (T^{3L}_{f}-2e_f \sin^2
  \theta_W)}{\sin \theta_W \cos \theta_W}[I_1(\tau_f, \lambda_f) - I_2
  (\tau_f, \lambda_f)], \text{where $f$ stands for $t$ or $b$ quark}
\nonumber \\ A_{H^{\pm}} &=& \frac{1-2\sin^2\theta_W}{\cos \theta_W
  \sin \theta_W} I_1(\tau_{H^{\pm}}, \lambda_{H^{\pm}})
\frac{m_W^2}{m_{H^{\pm}}^2}
\end{eqnarray}
and 
\beq
\tau_i = \frac{4 m_i^2}{m_h^2},~~~~~~ \lambda_i = \frac{4 m_i^2}{m_Z^2}
\eeq

The functions $I_1$ and $I_2$ are defined as

\begin{eqnarray}
I_1(a,b) &=& \frac{ab}{2(a-b)} + \frac{a^2 b^2}{2(a-b)^2}[f^0(a) -
  f^0(b)] + \frac{a^2 b}{(a-b)^2}[g(a) - g(b)], \nonumber \\ I_2(a,b)
&=& -\frac{ab}{2(a-b)}[f^0(a) - f^0(b)]
\end{eqnarray}

\beq g(\tau) = \left(\begin{array}{c} \sqrt{\tau - 1}\sin^{-1}
  (\sqrt{1/\tau)},~~~~~~~~~~~\text{if}~~\tau \ge 1,
  \\ -\frac{1}{2}\sqrt{1-\tau}[\text{ln}(\eta_{+}/\eta_{-}) - i
    \pi],~~\text{if}~~\tau <1
 \end{array}\right)
\eeq

$R_i^{h}$ in Equations~\ref{hgamgam} and \ref{hZgamma} denotes the scale
factor for $W$, top and bottom loop compared to their SM
values. Essentially, $R_W^{h}$ and $R_f^{h}$ are equivalent to the
scaling of $hVV$ and $hf\bar f$ coupling with respect to their SM
predictions. $R_{H^{\pm}}^h$ is the trilinear Higgs coupling
$g_{hH^+H^-}$ in all two-Higgs doublet models. The scale factors
$R_W^h$, $R_f^h$ and $R_{H^{\pm}}^h$ can be expressed in terms of the
model parameters, namely, $\tan \beta$ and $\sin (\beta - \alpha)$.  The
Higgs potential is general and is valid for all types of 2HDM.

\bigskip

\section{Decay amplitudes in models with triplet scalar(s)}
\label{app2}

\subsection{single triplet case}

\noindent
The decay amplitude for $h \rightarrow \gamma \gamma$ in models with single triplet is given by

\begin{eqnarray}
	\Gamma(h\rightarrow \gamma\gamma) &=& \frac{\alpha^2 G_F m_h^3}{128\sqrt{2}\pi^3} \bigg|
		\sum_f N_c Q_f^2 g_{hf\bar{f}}A^{\gamma\gamma}_{1/2}(\tau_h^f)
		+ g_{hW^+W^-} A^{\gamma\gamma}_1(\tau_h^W) \nonumber\\
		&& + \tilde{g}_{hH^\pm H^\mp} A^{\gamma\gamma}_0(\tau_h^{H^\pm})
		+ 4 \tilde{g}_{hH^{\pm\pm} H^{\mp\mp}} A^{\gamma\gamma}_0(\tau_h^{H^{\pm\pm}}) \bigg|^2\,,
\label{hgg}
\end{eqnarray}
 where  $\alpha$ is the fine-structure constant,
 $G_F$ is the Fermi coupling constant,
 $N_c=3 (1)$ for quarks (leptons),
 and $Q_f$ is the electric charge of the fermion in the loop.

The decay rate for $h\rightarrow Z\gamma$ in single-triplet case given by

\begin{eqnarray}
	\Gamma(h\to Z\gamma) &=& \frac{\alpha^2 m_h^3}{128\pi^3 v^2}
		\left(1-\frac{m_Z^2}{m_h^2}\right)^3 \bigg|
		\frac{1}{\sin \theta_W \cos \theta_W} \sum_f N_c Q_f (2I_3^f-4Q_f\sin^2 \theta_W) A_{1/2}^{Z\gamma}(\tau_h^f,\tau_Z^f)
		\nonumber\\
		&& + \cot\theta_W g_{hW^+W^-} A_1^{Z\gamma}(\tau_h^W,\tau_Z^W)
		- 2 g_{ZH^\pm H^\mp} \tilde{g}_{hH^\pm H^\mp}
		A_0^{Z\gamma}(\tau_h^{H^\pm},\tau_Z^{H^\pm}) \nonumber\\
		&& - 4 g_{ZH^{\pm\pm}H^{\mp\mp}} \tilde{g}_{hH^{\pm\pm}H^{\mp\mp}}
		A_0^{Z\gamma}(\tau_h^{H^{\pm\pm}},\tau_Z^{H^{\pm\pm}})\bigg|^2\,,
\label{hZg}
\end{eqnarray}
 where
 $\tau_h^i=4m_i^2/m_h^2$, $\tau_Z^i=4m_i^2/m_Z^2$ ($i=f,W,H^\pm,H^{\pm\pm}$),
 and $I_3^{f}$ are the third component of isospin of the fermion.
% In these equations the index $i$ runs over the various contributions to the loop.

\begin{eqnarray}
g_{ZH^\pm H^\mp} = -\tan \theta_W, \label{gz1} \\
g_{ZH^{\pm\pm}H^{\mp\mp}} = 2\cot 2 \theta_W 
\label{gz2}
\end{eqnarray}

The relevant loop functions are

%\section{\\Appendix B}
\begin{eqnarray}
	A_0^{\gamma\gamma} (x) &=& -x^2[x^{-1}-f(x^{-1})]\,, \nonumber\\
	A_{1/2}^{\gamma\gamma} (x) &=& 2x^2[x^{-1}+(x^{-1}-1)f(x^{-1})]\,, \nonumber\\
	A_1^{\gamma\gamma}(x) &=& -x^2[2x^{-2}+3x^{-1}+3(2x^{-1}-1)f(x^{-1})]\,, \nonumber\\
	A_0^{Z\gamma}(x,y) &=& I_1(x,y)\,, \nonumber \\
	A_{1/2}^{Z\gamma} (x,y) &=& I_1(x,y)-I_2(x,y)\,, \nonumber\\
	A_1^{Z\gamma}(x,y) &=& 4(3-\tan^2\theta_W)I_2(x,y)
		+ [(1+2x^{-1})\tan^2 \theta_W-(5+2x^{-1})]I_1(x,y)
\label{loops1}
\end{eqnarray}
 where
\begin{eqnarray}
	I_1(x,y) &=& \frac{x y}{2(x-y)} + \frac{x^2 y^2}{2(x-y)^2}[f(x^{-1})-f(y^{-1})]
		+ \frac{x^2 y}{(x-y)^2}[g(x^{-1})-g(y^{-1})]\,, \nonumber\\
	I_2(x,y) &=& - \frac{x y}{2(x-y)}[f(x^{-1})-f(y^{-1})]\;,
\label{loops2}
\end{eqnarray}
 with the functions $f(x)$ and $g(x)$ in the range $x<1$, given by
\begin{eqnarray}
	f(x) = (\sin^{-1}\sqrt{x})^2\,, \qquad
	g(x) = \sqrt{x^{-1}-1}(\sin^{-1}\sqrt{x})\,.
\label{loops3}
\end{eqnarray}  

\subsection{two-triplet case}
The decay width for the process $h \rightarrow \gamma \gamma$ in the two-triplet case is given as

\begin{eqnarray}
	\Gamma(h\rightarrow \gamma\gamma) &=& \frac{\alpha^2 G_F m_h^3}{128\sqrt{2}\pi^3} \bigg|
		\sum_f N_c Q_f^2 g_{hf\bar{f}}A^{\gamma\gamma}_{1/2}(\tau_h^f)
		+ g_{hW^+W^-} A^{\gamma\gamma}_1(\tau_h^W) \nonumber\\
		&& + \sum_{i=1,2}\left( \tilde{g_i}_{hH_i^\pm H_i^\mp} A^{\gamma\gamma}_0(\tau_h^{H_i^\pm})
		+ 4 \tilde{g_i}_{hH_i^{\pm\pm} H_i^{\mp\mp}} A^{\gamma\gamma}_0(\tau_h^{H_i^{\pm\pm}})\right) \bigg|^2\,,
\label{hgg2}
\end{eqnarray}

The decay rate for $h\rightarrow Z\gamma$ in the two-triplet case is given by

\begin{eqnarray}
	\Gamma(h\to Z\gamma) &=& \frac{\alpha^2 m_h^3}{128\pi^3 v^2}
		\left(1-\frac{m_Z^2}{m_h^2}\right)^3 \bigg|
		\frac{1}{\sin \theta_W \cos \theta_W} \sum_f N_c Q_f (2I_3^f-4Q_f\sin^2 \theta_W) A_{1/2}^{Z\gamma}(\tau_h^f,\tau_Z^f)
		\nonumber\\
		&& + \cot\theta_W g_{hW^+W^-} A_1^{Z\gamma}(\tau_h^W,\tau_Z^W)
		- \sum_{i=1,2}(2 g_{ZH_i^\pm H_i^\mp} \tilde{g_i}_{hH_i^\pm H_i^\mp}
		A_0^{Z\gamma}(\tau_h^{H_i^\pm},\tau_Z^{H_i^\pm}) \nonumber\\
		&& + 4 g_{ZH_i^{\pm\pm}H_i^{\mp\mp}} \tilde{g_i}_{hH_i^{\pm\pm}H_i^{\mp\mp}}
		A_0^{Z\gamma}(\tau_h^{H_i^{\pm\pm}},\tau_Z^{H_i^{\pm\pm}})) \bigg|^2\,,
\label{hZg2}
\end{eqnarray}

All the quantities in Equation\ref{hgg2} and \ref{hZg2} are given in
Equations~\ref{gz1}~-~\ref{loops3}. 

\section*{Acknowledgement}

We thank Satyaki Bhattacharya and Arpan Kar for useful discussions at various stages of the work.
This work was supported by funding available from the Department of Atomic  Energy, Government of India, for the Regional Centre for Accelerator-based Particle Physics (RECAPP), Harish-Chandra Research Institute. The authors acknowledge the hospitality of Indian Association for the Cultivation of Science, Kolkata, where a part of the study was carried out.

\bibliographystyle{JHEP}
\bibliography{paperbib}

\end{document}